\documentclass[12pt,preprint]{aastex}

\newcommand{\kms}{\rm km \, s^{-1}}
\newcommand{\ha}{H$\alpha$\ }

\newcommand{\mg}{$\rm Mg_2 \,$}
\newcommand{\mgb}{$\rm Mg \, \it b \,$}

\newcommand{\resn}{$10.5 \, \rm \AA$}
\newcommand{\resw}{$22.4 \, \rm \AA$}
\newcommand{\mgnobs}{$0.312 \pm 0.065 \,$}
\newcommand{\mgwobs}{$0.305 \pm 0.049 \,$}
\newcommand{\mgbcor}{$-0.001 \,$}
\newcommand{\mgnbcor}{$0.311 \pm 0.065 \,$}
\newcommand{\mgwbcor}{$0.304 \pm 0.049 \,$}
\newcommand{\mgap}{$0.03 \,$}
\newcommand{\mgxcor}{$0.05 \,$}
\newcommand{\mgnxcor}{$0.238 \pm 0.061 \,$}
\newcommand{\mgwxcor}{$0.232 \pm 0.040 \,$}
\newcommand{\mgno}{$0.237 \,$}
\newcommand{\mgwo}{$0.231 \,$}
\newcommand{\mgdif}{$0.006 \,$}
\newcommand{\mgadopt}{$0.238 \pm 0.061 \,$}

\newcommand{\mgstar}{$0.012 \pm 0.006 \,$}
\newcommand{\mgbstar}{$0.41 \pm 0.14 \,$}
\newcommand{\hbstar}{$3.63 \pm 0.35 \,$}

\newcommand{\EVI}{$2.020 \pm 0.082 \,$}
\newcommand{\EBV}{$1.555 \pm 0.098 \,$}

\newcommand{\opd}{1.688}
\newcommand{\dopd}{0.068}
\newcommand{\opdEBV}{$1.82 \pm 0.11 \,$}
\newcommand{\opdEVIccm}{1.84}
\newcommand{\opdEBVccm}{2.06}

\newcommand{\AV}{$4.67 \pm 0.19 \,$}
\newcommand{\AVEBV}{$5.01 \pm 0.32 \,$}
\newcommand{\AVdiffitz}{$0.3 \,$}
\newcommand{\AVdifccm}{$0.6 \,$}
\newcommand{\AVdav}{$4.7 \pm 0.8 \,$}
\newcommand{\AB}{$6.12 \pm 0.25 \,$}
\newcommand{\AI}{2.65}
\newcommand{\AIerr}{0.11}
\newcommand{\AKp}{$0.57 \pm 0.02 \,$}
\newcommand{\AVdifbmgas}{$0.4 \,$}
\newcommand{\AVdifbmcol}{$0.2 \,$}
\newcommand{\Acdif}{$0.6 \,$}

\newcommand{\RB}{3.623}
\newcommand{\RV}{2.765}
\newcommand{\RI}{1.569}
\newcommand{\RKp}{0.336}

\newcommand{\mBot}{$7.35 \pm 0.26 \,$}
\newcommand{\mVot}{$6.47 \pm 0.20 \,$}
\newcommand{\mIot}{$5.41 \pm 0.11 \,$}

\newcommand{\colVIe}{$1.100 \pm 0.065 \,$}
\newcommand{\colBVe}{$0.864 \pm 0.057 \,$}
\newcommand{\colBVt}{$0.88 \pm 0.13 \,$}
\newcommand{\colVIt}{$1.06 \pm 0.09 \,$}
\newcommand{\colBKt}{$3.85 \pm 0.27 \,$}

\newcommand{\Vd}{186.8}
\newcommand{\Vderr}{7.4}
\newcommand{\Vdap}{180.0}
\newcommand{\Vdaperr}{7.4}
\newcommand{\VdapDav}{$176 \pm 7 \,$}
\newcommand{\Vhel}{66.4}
\newcommand{\Vhelerr}{5.0}
\newcommand{\Vhelsedif}{1.9}

\newcommand{\SBI}{$19.01 \,$}
\newcommand{\SBIerr}{$0.11 \,$}
\newcommand{\logRe}{$0.460 \pm 0.051 \,$}
\newcommand{\FPD}{$2.92 \pm 0.37 \,$}

\newcommand{\SBFDdb}{$4.2 \,$}
\newcommand{\SBFDlt}{$4.0 \,$}
\newcommand{\SBFDagedb}{$3.4 \,$}
\newcommand{\SBFDagelt}{$3.0 \,$}
\newcommand{\DnB}{$366\arcsec \pm 77\arcsec \,$}
\newcommand{\DnBv}{$209 \pm 58 \,$}
\newcommand{\DnK}{$375\arcsec \pm 41\arcsec \,$}
\newcommand{\DnDB}{$3.08 \pm 0.85 \,$}
\newcommand{\DnDK}{$3.23 \pm 0.67 \,$}
\newcommand{\aveD}{$3.01 \pm 0.30 \,$}
\newcommand{\vGrp}{280}
\newcommand{\HubD}{$3.5 \,$}

\newcommand{\Do}{$23\farcm4 \pm 1\farcm2$}
\newcommand{\dmod}{$27.39 \pm 0.22 \,$}
\newcommand{\MB}{$-20.04 \pm 0.34 \,$}
\newcommand{\MV}{$-20.92 \pm 0.30 \,$}
\newcommand{\MI}{$-21.98 \pm 0.25 \,$}
\newcommand{\dmodCenA}{$28.05 \pm 0.11 \,$}
\newcommand{\MVCenA}{$-21.71 \,$}
\newcommand{\MVdif}{0.8}

\slugcomment{Accepted by The Astrophysical Journal on 2003 January 7}
\shorttitle{Fingerhut et al.}
\shortauthors{Extinction and Distance of Maffei 1}

\begin{document}

\title{The Extinction and Distance of Maffei 1}

\author{Robin L. Fingerhut\altaffilmark{1}, 
Marshall L. McCall\altaffilmark{1,2}, 
Michael De Robertis\altaffilmark{1},
Robin L. Kingsburgh\altaffilmark{1,2}, 
Michael Komljenovic\altaffilmark{1},
Henry Lee\altaffilmark{1}, 
and Ronald J. Buta\altaffilmark{3}
}

\altaffiltext{1}{
York University,
Department of Physics and Astronomy,
4700 Keele Street, Toronto, Ontario, Canada, M3J~1P3.
Email: rfinger@yorku.ca, mccall@yorku.ca
}

\altaffiltext{2}{
Visiting Astronomer, Observatorio Astron\'omico Nacional, Baja
California, M\'exico.
}

\altaffiltext{3}{
University of Alabama, 
Department of Physics and Astronomy, 
Tuscaloosa, Alabama, U.S.A., 35487--0324.
email: buta@sarah.astr.ua.edu
}

\begin{abstract}
We have obtained low- and high-resolution spectra 
of the core of the highly-reddened elliptical galaxy Maffei 1. 
From these data, 
we have obtained the first measurement of the \mg index, 
and have measured the velocity dispersion 
and radial velocity with improved accuracy.
To evaluate the extinction, a correlation between the
\mg index and effective $V-I$ colour
has been established for elliptical galaxies.
Using a new method for correcting for effective wavelength shifts,
the $V-I$ colour excess reveals that the optical depth 
of Galactic dust at $1 \, \rm \mu m$ is $\opd \pm \dopd$.  Thus,
$A_V=$ \AV mag,
which is lower by \AVdifbmgas mag
than previously thought.  
To establish the distance, the
Fundamental Plane for elliptical galaxies has been constructed
in $I$.  
The velocity dispersion of Maffei~1, measured to
be $\Vd \pm \Vderr \, \kms$, in combination with 
modern wide-field photometry in $I$, leads to a distance
of \FPD Mpc.
The $D_n$--$\sigma$ relation,
which is independently calibrated, gives \DnDB Mpc and \DnDK Mpc 
from photometry in $B$ and $K^\prime$,
respectively.
The weighted mean of the three estimates is \aveD Mpc,
which is lower than distances judged with
reference to M32 and the bulge of M31 from
the brightest stars seen at $K^\prime$.  Since the luminosity
of AGB stars at $K^\prime$ is strongly dependent upon age,
the lower distance suggests that the last epoch of star formation
in Maffei~1
occurred farther in the past than in these other systems.
The distance and luminosity make Maffei~1
the nearest giant elliptical galaxy.  
In the absence of extinction, the galaxy would be among
the brightest in the sky, and would have an apparent size
two-thirds that of the full Moon.
The radial velocity of Maffei~1 is $+\Vhel \pm \Vhelerr \, \kms$,
significantly higher than the accepted value of 
$-10 \, \rm \kms$.
The Hubble distance corresponding to the mean velocity
of Maffei~1, Maffei~2 and IC~342 is \HubD Mpc.
Thus, it is unlikely that Maffei~1 has had any influence on
Local Group dynamics.
\end{abstract}


\section{Introduction}
\label{intro}

Maffei~1 is a large
elliptical galaxy located in the middle of the
Zone of Avoidance ($l = 135.86$, $b = -0.55$).
Unfortunately, elucidation
of its fundamental properties, such as the absolute magnitude
and distance,
has been hampered by both its size and location.
Not only is it covered with myriads of foreground stars, 
but it also suffers from about 5 magnitudes of extinction in $V$
\citep{bm83}.  
The uncertainty in distance is especially
problematic given that it is positioned in the general
direction of the Andromeda galaxy.
Past work has led to the suggestion that the galaxy is near enough and
moving fast enough to have been in proximity to M31 in the
past 5 to 8 billion years \citep{mlm89,vbm93}.
If so, it could have influenced the
motions of galaxies in the Local Group, thereby invalidating the
binary hypothesis of Local Group timing \citep{kw59,lb81}.

Owing to the heavy obscuration,
the existence of Maffei~1
was unknown until 1968, when Paolo Maffei discovered it
on I-N and Ia-E plates acquired with the Schmidt telescope 
at Asiago Observatory during a search for T~Tauri stars and
diffuse nebulae
\citep{maf68}.
The subsequent
flurry of research activity led to its tentative identification
as an elliptical
galaxy \citep{sso71}, later 
confirmed by 
\citet{bm83,bm99} and \citet{lt93}.
By comparing the nuclear spectrum with spectra of
the spheroid of M31 and giant ellipticals like NGC~3379, 
\citet{sso71} estimated $A_V$ to be
$5.2 \pm 0.2 \, \rm mag$.  
Upon visual inspection
of a photographic spectrum, it was concluded
that the velocity
dispersion was ``not less than $150 \, \rm km \, sec^{-1}$ or
greater than $250 \, \rm km \, sec^{-1}$''; a
value of $200 \, \kms$ was adopted.  
Combined with a measurement of the core radius, 
an inferred total magnitude ($11.0 \, \rm mag$ in $V$) corrected for 
the extinction, 
and an assumed mass-to-light ratio, 
the distance was evaluated to be 1 Mpc.  Consequently,
it was suggested that Maffei~1 might be an unbound member
of the Local Group.  Certainly, the galaxy appeared to be
the nearest giant elliptical.

Subsequent estimates of extinction based upon studies of
foreground stars indicated that the result of \citet{sso71}
might be a magnitude
or more too high \citep[see][]{bm83}.
\citet{bm83} added two new measurements.
First, by measuring the $B-V$ colour of the inner region
via aperture photometry,
and comparing it with the mean colour of unobscured
ellipticals, $A_V$ was estimated to be
$5.3 \pm 0.4 \, \rm mag$.  
Also, from the total column density of gas in the direction of the
galaxy, judged from both HI and CO measurements, $A_V$
was estimated to be $4.9 \pm 0.4 \, \rm mag$.
Determinations based upon
analyses of foreground stars were clearly underestimates.
\citet{bm83} concluded that
$A_V = 5.1 \pm 0.2 \, \rm mag$.
Through a gross extrapolation of the aperture growth curve,
the apparent visual magnitude was determined to be $11.4 \, \rm mag$.
Combining the velocity dispersion estimated by \citet{sso71}
with the measurements of extinction and brightness,
a distance of 
$2.1^{+1.3}_{-0.8} \, \rm Mpc$ was derived from the relationship
between luminosity and central velocity dispersion observed in $V$
for elliptical galaxies, as defined by \citet{vo82}.
Thus, the galaxy no longer appeared to be in the Local Group,
but instead in a collection of galaxies known at the time as
the Ursa Major-Camelopardalis Cloud \citep{vau75}.

In an effort to derive a better estimate of the distance,
\citet{lt93} examined surface
brightness fluctuations in $K^\prime$ out to a radius of 
$75\arcsec$.
Using M32 and the spheroid of M31 as references, they
concluded that the 
distance was $4.15 \pm 0.5 \, \rm Mpc$.  In principle, this was the
best determination of distance yet.  However, there were
still reasons for caution, because of the possibility that
the measurement of the
fluctuation magnitude was compromised by contamination of
the field by faint foreground stars, globular clusters
in Maffei~1 itself, and variations in the galaxy-subtracted
background introduced by the known dust lanes.

An even more fundamental question is whether the mix of giants
in Maffei~1 responsible for the surface brightness
fluctuations at $K^\prime$ is similar to that in either of 
the reference fields.
One argument against this is that the fluctuation magnitude for
M32 was measured to be
brighter than that for M31 by $0.26 \pm 0.13 \, \rm mag$.  
Also, \citet{dav01} found that the
$K^\prime$ luminosity functions of AGB stars in M31 and M32
are intrinsically different, despite the agreement in the
magnitudes of their respective AGB tips.
Both \citet{lt93} and \citet{dav01} assumed
that M31 and M32 are equidistant.
If in fact the stellar mixes were identical,
then surface brightness fluctuations in $I$ would suggest that M32
is farther away than M31 by $0.15 \pm 0.07 \, \rm mag$ \citep{ton01}.
However, it has been demonstrated convincingly from studies of the
reddening of planetary nebulae in M32 that M32 is in front of
M31 \citep{for78,ric99}.  By studying the gravitational influence
of M32 on the disk of M31, \citet{byr76,byr78} concluded that
M32 is only $7.5 \, \rm kpc$, or $0.02 \, \rm mag$,
nearer than the nucleus of M31.
Thus, the discrepancy in the fluctuation magnitudes of these
two galaxies cannot be attributed to a distance effect alone.

Another problem with $K^\prime$ surface brightness fluctuations
in Maffei~1 is the observation that the
fluctuation magnitude
brightens by $0.46 \, \rm mag$
over the small range of radii studied 
(35\% of the length of the semimajor axis of the effective isophote in $I$:
\citealp{bm99}).
To account for this, \citet{lt93} assumed
``not without trepidation'' that the population of stars within
$24\arcsec$ was like that in M31 and that the mix from $24\arcsec$ to 
$75\arcsec$ was more like that in M32.
$B-R$ colours within the inner arcminute
of M31 and M32 
differ by $0.4 \, \rm mag$ \citep{pel93,wal87},  
supporting the argument that
different populations contribute to the surface brightness
fluctuations observed in these galaxies.
However,
in Maffei~1, the photometry of \citet{bm99} reveals
that both $B-V$ and $V-I$
change by only 0.13 mag
between the two
zones in question.
In other words, the stellar mix in Maffei~1 is relatively
homogeneous
throughout the observed region, while there is evidence that the reference 
field assigned to the inner observed zone differs in
stellar content from the reference field assigned to the
outer observed zone.
The origin of some
of the brightening in the fluctuation
magnitude with radius in Maffei~1
may actually be globular clusters, because \citet{bm02} have identified
12 candidates between $33\arcsec$ and $75\arcsec$ in images acquired
with HST.
The more reliable determination of distance, then,
might be the comparison of fluctuations within $24\arcsec$
of the nucleus with those observed
in the bulge of M31.  However, although the mean
absolute magnitude of $K^\prime$ fluctuations seen in Virgo
cluster ellipticals agrees with the value adopted by 
\citet{lt93} for the bulge of M31, the range for five galaxies
with total intrinsic $V-I$ colours like that of Maffei~1
is $1.09 \, \rm mag$
\citep{jen96}.

Using adaptive optics at CFHT, \citet{db01} were actually
able to measure $K^\prime$ magnitudes directly
for the brightest AGB stars in a $34\arcsec$ field located
$6\arcmin$ from the centre of Maffei~1.
Interpreting the stars as being similar to the most luminous
seen in the bulge of M31, they estimated a distance of
$4.4^{+0.6}_{-0.5} \, \rm Mpc$.  Although not entirely independent
of the result of \citet{lt93}, the technique has the advantage
that the luminosity of the brightest AGB stars does not appear to
vary much between M32 and the bulges of M31 and the Milky Way,
supporting the adoption of the AGB tip as a standard candle.
However, \citet{bre94} find a strong dependence of the
luminosity of AGB stars upon age. In \S~\ref{discussion},
we examine how much the age dependency may
extend the range in luminosity associated with the AGB tip,
and in turn, the effect on the distance to Maffei~1.

\citet{lt93} also applied the $D_n$--$\sigma$ relation to arrive at
an independent estimate of distance.  The result,
$4.2 \pm 1.1 \, \rm Mpc$, agreed
with that from surface brightness fluctuations, but the
observational ingredients were highly uncertain, being
founded upon their surface photometry in $K^\prime$ 
(to only $90\arcsec$, i.e., 30\% of $D_n$), 
an average $B-K$ colour for ellipticals,
and the velocity dispersion of
\citet{sso71}.  The method is re-examined
in \S~\ref{distance} using modern data.

Distances derived from stars within Maffei~1 do not directly
give information about the luminosity of the galaxy without
good integrated photometry.  Realizing that adequate photometry
was lacking, \citet{bm99} acquired images of
Maffei~1 in $B$, $V$, and Cousins $I$ using
a CCD camera attached to the Burrell Schmidt telescope at
Kitt Peak.  After carefully removing foreground stars,
the apparent $V$ magnitude was found to be $11.14 \pm 0.06 \, \rm mag$,
confirming the earlier results of 
\citet{sso71} and \citet{bm83}.
Using the lower distance of \citet{bm83},
the absolute magnitude ($M_V = -20.6$) would make Maffei~1
an intermediate elliptical by the standards of galaxies in
the Virgo cluster.  Using the larger distance of \citet{lt93},
Maffei~1 would have a luminosity ($M_V = -22.1$) 
comparable to that of the
brightest members of the cluster.

It is clear that there is a need to gauge the distance to Maffei~1 
in a new way.  As it turns out, it is also necessary
to re-evaluate the extinction.  Since 1983, most authors have
adopted the extinction to be that recommended by \citet{bm83}.
It is behind all of the recent
distance determinations described above.
The map of extinction derived
by \citet{sfd98}
from direct observations of dust emission can not be applied
because it is uncalibrated at Galactic latitudes below $5^\circ$.
In fact, nobody
has actually measured the {\it extinction} of Maffei~1.
Rather, any evaluation of extinction has come from a measurement
of the {\it reddening}, in combination with a reddening law
and an assumption about the ratio of total
to selective extinction, $R_V = A_V / E(B-V)$.  

It is well known that,
as a consequence of shifts in effective wavelengths, the
ratio of total to selective extinction for a star depends upon both the
spectral energy distribution (SED) and the reddening.
Until recently, the subject had never been explored for galaxies,
and even for stars it had really only been investigated thoroughly
for $B$ and $V$.
\citet{ma00} and \citet{mlm02} have shown that 
the ratio of total to selective extinction for galaxies does
indeed vary with morphological type, reddening, and, especially,
redshift.  In fact, $R_V$ for a
star-forming galaxy is different from that for a star
of the same colour.

In evaluating the extinction in a particular
passband, most researchers use a colour excess, traditionally
$E(B-V)$, and a reddening law, which gives the extinction as
a function of wavelength, usually normalized to $E(B-V)$.
There are two problems with this approach.  First, the object
used to gauge the colour excess does not always have the
same spectrum as the object for which the extinction is desired.
Second, objects used to define the reddening law may be different
altogether.  Even a so-called monochromatic reddening law may
be normalized to a colour excess defined by broad-band filters,
so the coefficients of the law remain object-dependent.
Thus, the derivation of an extinction of a source may be filled
with inconsistencies.  For example, a typical researcher might
use the map of \citet{sfd98} to gauge the extinction of 
a spiral galaxy.  The maps are calibrated using estimates of
$E(B-V)$ for elliptical galaxies spread all over the sky
(with intrinsic colours judged from the \mg index).
The conversion of colour excesses to extinction might be accomplished
using the reddening law
of \citet{ccm89}, which specifically is founded
upon studies of O and B stars.  In the end, a colour index
for elliptical galaxies is used with a reddening law for O and
B stars to evaluate the extinction of a spiral galaxy!

As mentioned previously,
\citet{bm83} gauged the colour excess of Maffei~1 by
two methods.  The colour excess judged from the comparison
of the apparent $B-V$ colour of the galaxy with the
mean colour for ellipticals was converted
into an extinction using $R_V = 3.5$, which
was estimated from the intrinsic colour
and colour excess via the relation derived by
\citet{ols75} for stars.
The colour excess judged from
the column density of gas along the line of sight was founded
upon a correlation observed for O and
B stars in the Milky Way.  In this case, $A_V$ was computed
using $R_V = 3.2$,
appropriate for those stars \citep{ols75}.  
This second estimate of $A_V$
really was applicable only to O and B stars,
not to Maffei~1, or even to stars with the same colour as
Maffei~1.  

For objects at high galactic latitudes, a small error in $R_V$
has rather insignificant consequences (except for studies of
large-scale flows: see \citealt{hud99}).  However, the error
is amplified for heavily-obscured objects like
Maffei~1.
Properly taking into account effective wavelength shifts
\citep{mlm02}, $R_V$ is 3.22
for an elliptical galaxy and 3.00 for a  B0~V star,
given that both are at rest
and obscured
comparably to Maffei~1.  If one uses the colour excesses
to solve for the optical depth along the line of sight
at 1 $\micron$ (see \S~\ref{s_exmaf1}),
the colour of Maffei~1 leads to an extinction
of $4.9 \, \rm mag$, which is $0.4 \, \rm mag$ lower than
estimated by \citet{bm83}, and
the column density of gas gives an extinction of $4.5 \, \rm mag$,
also $0.4 \, \rm mag$ lower.

An additional reddening estimate was recently obtained
by \citet{dav02}, who measured the mean $H-K$
colour of bright giants in Maffei~1 and derived
$E(H-K) \rm = 0.28 \pm 0.05 \, mag$ based on the assumption
that the brightest stars have the same intrinsic
colours as M giants in Baade's Window. Following the
procedure described in \S~\ref{s_exmaf1},
this colour excess leads to an extinction of
$\rm 4.7 \pm 0.8 \, mag$ in $V$.
In this paper, the extinction of Maffei~1 is estimated with
improved accuracy by measuring the 
\mg index for the first time,
and by deriving a value for the ratio of
total to selective extinction consistent with the SED, 
extinction, and redshift.
Then, the distance is determined for the first time from the
Fundamental Plane using a modern measurement of 
the velocity dispersion in combination with the effective radius
determined by \citet{bm99} from wide-field surface photometry.
A new determination of the radial velocity is used to evaluate
the degree to which Maffei~1 departs from the Hubble flow
and the likelihood of any interaction with the Local Group
subsequent to the Big Bang.

Observations are presented in
\S~\ref{observations}, and reductions are outlined in
\S~\ref{reductions}.  Measurements of the \mg index, velocity
dispersion, and radial velocity are described in \S~\ref{measurements}.
The extinction is determined in \S~\ref{extinction}, and the
distance is derived in \S~\ref{distance}.
Implications for the Local Group are discussed in \S~\ref{discussion}.
Finally, the properties of Maffei~1 are summarized in
\S~\ref{summary}.

\section{Observations}
\label{observations}

All spectroscopy of Maffei~1 was carried out with the
Boller and Chivens long-slit spectrograph attached to the $f/7.5$ focus
of the $\rm 2.12 \, m$ telescope at the Observatorio
Astron\'{o}mico Nacional in San Pedro M\'artir, Baja California,
M\'exico.  
The detector employed was a Tektronix CCD coated with Metachrome II,
which consisted of $\rm 24 \, \mu m$ square pixels arranged in a 
$1024 \times 1024$ array.  The readout noise was
$3.0 \, \rm e^{-} \, pix^{-1}$, and the dark current was a negligible
$0.76 \, \rm e^{-} \, pix^{-1} hr^{-1}$.
The readout electronics did not permit overclocking the CCD, so frames
did not include overscan pixels.  Consequently, to monitor
the bias level, a bias exposure was taken after
each target exposure.
For all observations of Maffei~1,  the slit
was oriented at a position angle of
$170^\circ$, which was close to the atmospheric dispersion direction.
This was only $4^\circ$ away from the minor axis of
the galaxy \citep{bm99}.

Low resolution spectra suitable for measuring the
\mg index were obtained on 1994 December 1 UT.
They were acquired in first order
with a $300 \, \rm lines/mm$ grating blazed at $5000 \, \rm \AA$,
which yielded coverage from 3450 to $7550 \, \rm \AA$ at
a dispersion of $4.00 \, \rm \AA \, pix^{-1}$.
To test the sensitivity of results to the aperture size,
spectra were acquired with two different slit widths,
$220 \, \rm \mu m$ ($2\farcs9$) and $600 \, \rm \mu m$ ($7\farcs8$).
The effective resolution, determined from measurements of the 
night sky line
$\rm [O~I]\lambda 5577$, was 2.6 pix FWHM (\resn) for the
narrow-slit spectrum and 5.6 pix FWHM (\resw) for the
wide-slit spectrum.
The useful length of the slit was 250 pix ($3\farcm9$).
The exposures of Maffei 1 were each 1800 seconds long.
Spectra of a helium-argon arc lamp were taken at the beginning
and end of the night to measure the wavelength scale.
The spectrophotometric
standards HD~217086, Hiltner~102, HD~84937 and G191B2B
were observed with a 
$600 \, \rm \mu m$ slit to calibrate fluxes.
To correct the data for any bias pattern and for pixel-to-pixel
variations in response,
nine bias exposures and three dome flat field exposures
were acquired at both the beginning and end of the night.
A twilight flat field exposure was taken at the end of the night
to determine the slit illumination function.

Spectra at high resolution appropriate for measuring the
velocity dispersion (and radial velocity)
were obtained on 1994 December 2 UT.
They were obtained in second order
with a $1200 \, \rm lines/mm$ grating blazed at $10,000 \, \rm \AA$,
which yielded a dispersion of $0.50 \, \rm \AA \, pix^{-1}$.
The tilt was set so that spectra spanned 5120 to 5630 $\rm \AA$.
A BG39 order-blocking filter was used to exclude light 
from first order.
The slit width was set to $270 \, \rm \mu m$ (3\farcs5), which
mapped onto $1.4 \, \rm pix$ ($0.7 \, \rm \AA$).
In other words, the slit was undersampled.  Measurements of
night sky $\rm [O~I]\lambda 5577$ revealed that the effective
resolution on average was $2.03 \, \rm pix$ FWHM ($1.05 \, \rm \AA$). 
To improve the signal-to-noise ratio, pixels {\it along the slit}
were on-chip binned in groups of 4.

High-resolution
observations were made in a manner designed to maximize velocity
precision.  Seven 1200-second exposures
of Maffei~1 were acquired, each
bracketed by observations of a helium-argon arc lamp.
Template spectra were provided
by four G and K giants with known radial velocities and with
positions in the sky comparable to that of Maffei~1, 
all selected from
the Bright Star Catalog Supplement \citep{hof83}.
Relevant data are summarized in Table~\ref{tabkin}.
The stars were observed, bracketed by arc exposures,
before, in between, and just after
the Maffei~1 exposures, with exposure times ranging 
from 30 to 40 seconds.
One giant was observed twice to check repeatability.
Since the orientation of the slit was almost north-south
(the direction of atmospheric dispersion for Maffei~1),
errors in tracking and guiding would have moved the stars
perpendicular to the slit, in essence helping to fill the
slit comparably to the nucleus of Maffei~1.
After completing the observations, five dome flat field
exposures were taken.  No twilight flat was acquired.
No flux standards were observed.

To compare the properties of Maffei~1 with those of more distant galaxies,
it is necessary to correct measurements
to a common physical scale.  Thus, the slit width and the
length of extraction apertures must be converted to angular units.
From the telescope parameters, the scale at the slit was
computed to be $\rm 12\farcs97 \, mm^{-1}$.
The scale along the spatial direction at the detector had to
be measured.  A spectrum of a field in the galaxy
UGC~4115, which was acquired on 1994 November 30 UT with
the same low-resolution configuration as used for Maffei~1, 
was suitable for this purpose as it included an HST~GSC star and
compact HII region of known separation \citep{km98}.
Also, the slit used for the low-resolution observations
of Maffei~1 the following night admitted the light from
four stars which were well separated from the core of
the galaxy and whose positions were known precisely in $V$
\citep{bm99}. From these data, the scale along the slit was
determined to be $0\farcs925 \rm \, pix^{-1}$.

\section{Reductions}
\label{reductions}

Reductions were carried out completely within
IRAF\footnote{ IRAF
is distributed by the National Optical Astronomy Observatories, which
is operated by the Association of Universities for Research in
Astronomy, Inc., under contract to the National Science Foundation.}.
Each night was reduced independently.
The mean bias level of each image of a source was judged from the
bias frames taken closest in time, and then subtracted.
Compensation for pixel-to-pixel variations in the bias was achieved by
subtracting a zero-level correction frame constructed
by averaging all of the median-subtracted bias images taken during the night.
To correct for pixel-to-pixel variations
in response, a master dome flat was created by
averaging all bias-corrected dome flat-field images.

Subsequent reductions of the low resolution spectra proceeded as
follows.
First, the master dome flat was rectified.
The twilight flat was used to
derive a smooth illumination function describing how the
response varied along the slit.  Flat field corrections were
accomplished by dividing the product of the master dome
flat and the illumination function into each source
image.
Dispersion solutions in two dimensions
were separately derived for the arc
spectra taken at the beginning and end of the night, and then
averaged using FITCOORDS.
Curvature along the
dispersion direction was traced using
spectra of standard stars and extragalactic HII regions
displaying a strong continuum.
Finally, spectra of targets were transformed
to linear wavelength and spatial scales
via TRANSFORM.
One-dimensional versions of the target spectra were extracted
using APALL.

Each one-dimensional spectrum of Maffei~1 was created
by summing without weights the 20 rows ($18\farcs5$)
in which the signal was noticeably above the background.
The background signal was determined using 
sampling windows on either side,
20 rows ($18\farcs5$) each in width.
The innermost boundary of each window was located
30 rows ($27\farcs8$) from the midpoint of the galaxy signal. 
The proximity of these windows to the centre of the galaxy
was necessitated by the complexity of the slit profile.
A linear fit to the median signal within each background window
was used to produce the background-subtracted spectra.
Based upon the surface brightness profile of \citet{bm99},
the flux from the galaxy contributing
to the signal in the background windows
is 5.9\% of that at the centre.
In \S~\ref{mg2}, we evaluate the degree to which this
effects our measurement of the \mg index and derive
an appropriate correction.

The sensitivity function was computed by simultaneously fitting
the spectra of HD~217086, Hiltner~102 and HD~84937.
The spectrum of G191B2B was excluded owing to anomalous residuals.
Then, the spectra of Maffei~1 were converted to a flux scale with
the aid of CALIBRATE.
Mean extinction coefficients for San Pedro 
Martir were employed throughout
\citep{sch97}.

Reductions of the 
high resolution spectra after pre-processing
proceeded somewhat differently.
The master dome flat
was flat enough to divide directly into the source spectra
without it having to be rectified.
Dispersion solutions for each of the five template spectra and
each of the seven spectra of Maffei~1 were derived from
the HeAr arc spectra bracketing them.  Each spectrum was
traced individually, and then, with the appropriate dispersion
solution, transformed to linear spatial and wavelength scales.  
Fits to the 
night sky line $\rm [O~I]\lambda 5577$ 
revealed outstanding stability in both wavelength
and focus.  Over the seven spectra of Maffei~1, 
the wavelength of the line
varied by only $\pm 0.05 \, \rm \AA$ ($\pm 3 \, \kms$), and
the full width at half maximum was stable to $\pm 0.02 \, \rm \AA$
($\pm 1 \, \kms$).
Spectra of Maffei~1 and the template stars were shifted 
(by up to $0.13 \, \rm \AA$)
to force the wavelength of the [O~I] line to a common 
value, specifically $5577.35 \, \rm \AA$
\citep{bow60}.
Then, one-dimensional versions were extracted using APALL
by averaging with signal-dependent weights
the 8 rows ($29\farcs6$)
centered on the peak flux from the source.
The choice of window was driven by the extent of 
detectable flux in the Maffei~1 spectra.
The background was defined by two windows roughly $1\arcmin$
wide centered approximately $65\arcsec$ away on either side
of the nucleus of the galaxy.
Finally,
an unweighted average of the seven one-dimensional
spectra of Maffei~1 was computed to arrive at the
spectrum used for measuring the velocity
dispersion and radial velocity.  This spectrum is compared
with that of one of the template stars in 
Figure~\ref{fighispec}.  The signal-to-noise ratio of
the Maffei~1 spectrum  is about 25 per pixel.

\section{Measurements}
\label{measurements}

\subsection{\mg Index}
\label{mg2}

The \mg index is a measure of the deficit 
of flux at the MgH and Mg $b$ absorption features 
relative to the continuum level defined by neighbouring 
spectral regions
\citep[see \S 2 of][]{wor94}.
For unobscured elliptical galaxies, it is
correlated with colour (see \S~\ref{s_colmg}).  Consequently, it
can be used to judge the intrinsic colour of a reddened
elliptical.

The low-resolution spectrum of Maffei~1 from 
which the \mg index was determined is shown in 
Figure~\ref{figlospec} with the feature and continuum 
sampling bands marked.
In measuring the index, we
followed \citet{hud99} and adopted
the rest-frame values defining the feature and continuum bands
to be those recommended
by \citet{bfg84}.
This ensures consistency with the \mg indices used to 
construct the colour--\mg relation from which we 
determine the reddening (\S~\ref{s_colmg}). 
Measurements were made using IRAF's SPLOT module, which
computes the equivalent width of a spectral feature
by taking the sum of the continuum-subtracted 
pixels between the two wavelengths defining the feature.
The continuum was determined by drawing a straight line
defined by the wavelength midpoints of the surrounding
continuum bands and the mean flux within each continuum band.
The continuum level at the feature was taken to be
the interpolated flux at the midpoint of the feature bandpass.
The error in the continuum level was computed as the average
standard error of the mean flux measured within
the blue and red continuum bands.
From the wide-slit spectrum,
\mg = \mgwobs mag,
and from the narrow-slit spectrum,
\mg = \mgnobs mag.
The uncertainty in each measurement is
computed from the standard error in the
equivalent width, for which the error in
the line flux is the average of the 3$\sigma$
errors in the fluxes measured in the blue and red continuum
bands and the error in the continuum level
is as described above.

Significant variations in the focus with wavelength
can change how light is distributed perpendicular to
the dispersion, thereby adding uncertainty to
measurements of the \mg index at small radii
\citep[see][]{ffi95}.
To determine the extent to which our measurements
suffer from this effect,
we examined the FWHM as a function of wavelength of
the spectrum of a foreground star
on the images containing the Maffei~1 spectra.
The width of the star remained constant to within 0.2 pix ($0\farcs2$)
over the wavelength range encompassing the \mg feature
and the two surrounding bands used to define the continuum level.
Since the spectra were produced by integrating the
light from the galaxy over 20 pixels, the degree to which 
focus instability contributes
to the uncertainty in our \mg measurements
is effectively negligible.

Our observations of the spectroscopic standard HD~84937
can be used to judge how close our measurements of the \mg
index of Maffei~1 are to the Lick-IDS system, on which
measurements of the reference galaxies are placed.
\citet{bfg84} normalized all galaxy
spectra by scans of a quartz-iodide tungsten lamp,
but did not place them on a flux scale.
Our low-resolution spectrum of HD~84937 with the
highest signal-to-noise was
obtained on the night of 1994 December 2 UT.
We used this spectrum to measure
the \mg index as well as the equivalent widths of
the \mgb and $\rm H \beta$ features.
The measurements were made using the same procedures
adopted by \citet{wor94}.
The observed resolution
($\rm 10.1 \, \AA$) was comparable to the value of
$\rm 8 \, \AA$ reported by \citet{wor94}.
Our results are \mgstar mag for the \mg index,
\mgbstar $\rm \AA$ for the equivalent width of the
\mgb feature, and \hbstar $\rm \AA$
for the equivalent width of $\rm H \beta$.
In comparison, the Lick-IDS values obtained by \citet{wor94}
are $\rm 0.024 \pm 0.008 \, mag, \, 0.50 \pm 0.23 \, \AA, \,
and \, 3.49 \pm 0.22 \, \AA$ for the \mg, \mgb and
$\rm H \beta$ features, respectively.
We conclude that our measurements can be considered
to be consistent with the Lick-IDS system, despite
the differing approaches to flux calibration.

From the $V$-band surface brightness profile of \citet{bm99},
the surface brightness of the galaxy at the radius of the
background sampling windows has dropped to 3.5\% of
that in the centre.
The degree to which this effects our measurement of the
\mg index can be estimated from
\begin{equation}
\label{eqskycor}
\rm \left(Mg_2\right)_0 = \left(Mg_2\right)_{obs} + 2.5 \log 
\left[\left(1-\it{k_l}\right) / \left(1-\it{k_c}\right)\right] 
\end{equation}
where $\rm (Mg_2)_{obs}$ is the measured \mg index,
$\rm (Mg_2)_0$ is the \mg index which would be measured if there
were no galaxy signal in the background regions,
and $k_l$ and $k_c$ are the ratios of the surface brightness
of the galaxy
at the midpoint of the background regions
to the surface brightness at the centre of the galaxy
in the line and continuum, respectively.
If the strength of
the \mg feature did not vary with radius, the observed
\mg index would be equivalent to the true \mg index.
However, gradients in \mg are known to exist;
\citet{dsp93} found the mean of the radial gradients
in \mg for ellipticals to be -0.059 mag per normalized
radius $r/r_e^*$, where $r$ is the radius at which the
\mg index is measured and $r_e^*$ is related to the effective
radius $r_e$ of the circular aperture encompassing half of
the total light of the galaxy in $B$.
We use this result to find the expected change in the \mg
index from the core of Maffei~1 out to the midpoint of the
background windows ($37\arcsec$).
For line indices measured along the minor axis, as is the
case for our \mg measurement, the normalization factor,
$r_e^*$, is given by 
$r_e(1-\epsilon)^{1/2}$, where $\epsilon$ is the ellipticity.
Since $r_e = 222\arcsec$ and $\epsilon = 0.271$ \citep{bm99},
then $r/r_e^*=0.19$.
The change in the \mg index from the centre of
Maffei~1 to $r/r_e^*$ is therefore expected to be -0.011~mag.
We compute $k_l/k_c=1.011$
given that the change in the \mg index is
equal to $-2.5 \log (k_l/k_c)$.
From the $V$-band surface brightness profile of \citet{bm99},
$k_c=0.059$, so $k_l=0.060$. By Equation~\ref{eqskycor},
the magnitude of the
correction to the observed \mg index is therefore \mgbcor mag.
The corrected indices are \mgwbcor mag and
\mgnbcor mag for the wide- and narrow-slit spectrum, respectively.

The centers of the two continuum bands are $407 \rm \, \AA$
apart. Thus, for heavily reddened galaxies, the \mg index
can be affected by differential extinction. In the case of
Maffei~1, the blueward band is extinguished by roughly \Acdif mag
more than the redward band
(compare Figure~\ref{figlospec}a with Figure~\ref{figlospec}b).
In \S~\ref{s_mgmaf1}, we employ an iterative process
to arrive at the \mg index that would be measured if Maffei~1
were reddened comparably to the reference galaxies defining
the $(V-I)$--\mg relation.

\subsection{Radial Velocity and Velocity Dispersion}
\label{vdisp}

The radial velocity of Maffei 1 was measured by
cross-correlating the spectrum of the galaxy
with each of the five template spectra.  
This was accomplished
using the task XCSAO in the package RVSAO
of IRAF \citep[see][]{td79}.
To avoid
contamination by night sky $\rm [O~I] \lambda 5577$,
and to improve the signal-to-noise ratio of the cross-correlation
peak, analyses were restricted to the wavelength range
$5125$--$5570 \, \rm \AA$, which is where absorption lines are
strongest and most numerous.

The results for the
heliocentric
velocities 
are given in column 5 of Table~\ref{tabkin}.
The sign of the velocities was confirmed by visual
inspection.
The errors in column 6 are those generated by XCSAO according to
the algorithm derived in \citet{td79}.
The result from HD~27224 is clearly deviant with respect
to the other four measurements, so it is rejected.
We suspect that the radial velocity of this star is errant.  
Thus, the heliocentric velocity of Maffei~1
is adopted to be the unweighted average of the other
four values.  Their outstanding agreement confirms that
offsets caused by variations in the filling of the slit
are small.
The uncertainty in the radial velocity
of Maffei~1 is adopted to be the standard deviation
of the different determinations, excluding HD~27224,
as opposed to the least-squared error (i.e., derived
from the inverse sum of the squares of the uncertainties
in the individual measurements). 
Since the standard deviation encompasses the
uncertainty in the radial velocities of the template stars
as well as the XCSAO fitting uncertainty, whereas the least-squared
error only reflects the latter,
it is likely that the standard deviation is a more
realistic estimate of the uncertainty.
We note that the least-squared error is
only $\Vhelsedif \, \kms$ larger than the adopted error.

The heliocentric radial
velocity of Maffei~1 is $+\Vhel \pm \Vhelerr \, \kms$,
which is significantly higher than
previously thought.  For their simulations of the dynamical
evolution of the Local Group, \citet{vbm93}
adopted a value of $-10 \, \rm km \, s^{-1}$, which was the 
``mean'' tabulated by \citet{huc86}.
Its origin was a photographic measurement by 
\citet{sso71}, who estimated the uncertainty
to be $50 \, \rm km \, s^{-1}$.
\citet{bm99} quoted a velocity of $-87 \, \rm km \, s^{-1}$,
which was derived from preliminary versions of the spectra
presented here and analyzed using the Fourier quotient technique
\citep{kor88}.
However, the sign was incorrectly established, and the magnitude
was affected by an inferior wavelength calibration.
Clearly, the new result has significant implications for 
dynamical timing, and especially conclusions about the 
relevance of Maffei~1 to motions in the Local Group.

The velocity dispersion of Maffei~1 was also measured
by cross-correlating the spectrum of the galaxy with each
template spectrum using the XCSAO task.
As was the case for the determination of the radial velocity,
the spectral range was restricted to the 890 pixels
in the range $5125$--$5570 \, \rm \AA$, particularly because 
the overwhelming
majority of the ``power'' in absorption lines was blueward of
$\rm [O~I]\lambda5577$.  Experiments showed that both wider
and narrower ranges led to more scatter in the results
(although the mean hardly changed).
The continuum was removed from all spectra before processing.
Ends of the spectra were windowed with a 5\% cosine bell.
Changes in the width of this bell had little effect on the results.
Low and high frequency noise were filtered using a cosine bell
with a cut-on from 5 to 20 and a cut-off from 150 to
400. The latter two values were determined graphically
from the power spectrum.
Of the four values,
only the last had any significant impact
upon the results, and then only if a wildly discrepant choice
were made. 

The results for the velocity dispersions
are given in column 7 of Table~\ref{tabkin}.
All determinations are in excellent agreement.
Thus, the velocity dispersion of Maffei~1
is adopted to be their unweighted average, namely
$\Vd \pm \Vderr \, \kms$.
\citet{td81} found that when the velocity dispersion is much
larger than the dispersion in the line-spread function,
the uncertainty in the velocity dispersion is
$1.2 \delta v_\sun$, where $\delta v_\sun$ is
the uncertainty in the heliocentric radial velocity.
We adopted this result to compute the
uncertainties in the velocity dispersions given in
column 8 of Table~\ref{tabkin}.
Conservatively, we adopt the uncertainty in the
velocity dispersion of Maffei~1
to be the least-squared error.

Interestingly, the estimate of \citet{sso71} was close to the
right answer.
At the time, this
led to the conclusion that Maffei~1 was located at the edge
of the Local Group.  However, it will be shown that 
our knowledge of the Fundamental Plane today leads to a very
different conclusion.

\citet{sso71} also
noted a discontinuity in spectral lines through the nucleus,
and estimated a rotational velocity of $130 \, \rm km \, s^{-1}$
averaged over $30\arcsec$.
The position angle of the slit employed was not identified,
but it might be presumed to have been along the major axis.
There is no evidence of a discontinuity in the two-dimensional
version of our combined spectrum, which was taken along the
minor axis, although it is possible that on-chip binning
reduced the spatial resolution enough to hide it.

\subsection{Aperture Corrections}
\label{apcor}

Both the \mg index and the velocity dispersion vary
across the face of an elliptical galaxy.  Thus,
to compare measurements for Maffei~1
with those for other galaxies,
as must be done to derive the reddening and distance,
the measured values must be adjusted to a common standard aperture.

\citet{jfk95} derived an 
empirical aperture correction formula for velocity
dispersions as a function 
of the equivalent angular radius 
$\rm \theta_{obs}$ of an observing aperture 
normalized to the angular radius $\rm \theta_e$ 
of the effective aperture of the galaxy, i.e., the circular aperture
enclosing half of the total light in $B$.
While their adopted power-law fit is adequate for typical galaxies,
it deviates considerably from their empirical 
data in the region $\log (\rm \theta_{obs} / \theta_e) \leq -1.4$,
which is where a nearby giant object like Maffei~1 resides. 
A better fit to the data points in this region is given by
\begin{equation}
\label{eqapcor}
\rm \log \left( \sigma_{e8} / \sigma_{obs} \right) = 0.032 \log 
\left[ \theta_{obs} / (\theta_e / 8) \right]
\end{equation}
where $\rm \sigma_{e8}$ is the 
``standard''  velocity dispersion which would 
be measured through an aperture with a radius of one eighth of 
the radius of the effective aperture of the galaxy,
$\rm \sigma_{obs}$ is the observed velocity dispersion 
measured through an aperture of width $x$ and length $y$, and
$\rm \theta_{obs}$, the equivalent aperture radius, is given by
$1.025 ( x y / \pi )^{1/2}$.
\citet{dsp93} found that the radial gradient in 
\mg is similar to that in the velocity dispersion, 
so equation~\ref{eqapcor} can be used to correct an \mg index,
too, simply be replacing $\log \sigma$ with \mg.

The radius of the effective aperture of Maffei~1 
in $B$ is $222\arcsec$ \citep{bm99}.
Based upon the angular scales along and perpendicular
to the slit (\S~\ref{observations}), the aperture dimensions for 
the two low-resolution spectra
and the high-resolution spectrum are 
$2\farcs85 \, \times \, 18\farcs50$,
$7\farcs78 \, \times \, 18\farcs50$, and
$3\farcs50 \, \times \, 29\farcs60$, respectively.
The equivalent aperture radii for the two low-resolution spectra
and the high-resolution spectrum are therefore
$4\farcs20$, $6\farcs94$,
and $5\farcs89$, respectively.
The aperture-corrected velocity dispersion
is $\Vdap \pm \Vdaperr \, \kms$,
with all of the uncertainty coming from the measurement
of the velocity dispersion itself, owing to the small size of the
aperture correction.
For the \mg indices, the aperture corrections for both the
wide- and narrow-slit spectra amount to \mgap mag.
We note that this is consistent with the change in
the \mg index implied by the small change in $V-I$
between the equivalent aperture radii and $\theta_e$.
Based on the
observations of \citet{bm99}, the changes in $V-I$
from $4\farcs20$ to $\theta_e$ (narrow-slit) and
$6\farcs94$ to $\theta_e$ (wide-slit)
imply changes in the \mg index of $0.036 \pm 0.006 \, \rm mag$
and $0.032 \pm 0.006 \, \rm mag$, respectively, according to
the $(V-I)$--\mg relation given by equation~\ref{eqcolmg}
(see \S~\ref{s_colmg}).
The agreement between the aperture corrections derived from
equation~\ref{eqapcor} and the declines in \mg predicted from the
changes in $V-I$ confirms the applicability of our
correction scheme based on the empirical correction data
of \citet{jfk95}.
Since the reference \mg indices were aperture-corrected
using the data of \citet{jfk95},
we adopt for consistency the aperture correction given
by equation~\ref{eqapcor} for Maffei~1, namely \mgap mag.

\section{Extinction}
\label{extinction}

\subsection{The $(V-I)$--\mg Relation}
\label{s_colmg}

\citet{ben93}
showed that for all dynamically hot systems,
the $B-V$ colour is tightly correlated with the \mg index.
Any scatter in the relationship appears to be random, i.e.,
not tied to another parameter.
In principle, then, it is a useful tool for evaluating the intrinsic
colour of an obscured galaxy, and, in turn, the reddening,
as was done by \citet{sfd98}.

Unfortunately, owing to the heavy obscuration, the properties
of Maffei~1 in $B$ remain considerably more uncertain than
in $V$ or $I$.  It is productive, therefore, to seek out a
relationship between $V-I$ and the \mg index.
\citet{hud99} summarizes a large set of measurements 
of the \mg index for elliptical galaxies, including
the extensive collection of \citet{fab89},
but all homogenized to the system of \citet{bfg84},
which was developed at Lick Observatory with the
Lick-Image Dissector Scanner (IDS).
For 97 of the galaxies,
effective and total photoelectric $V-I$ colours have been
determined by \citet{bw95}.
The effective $V-I$ colour, defined as the 
integrated colour within the circular aperture 
transmitting half the total blue light, 
is preferred over the total colour, as the latter 
is derived by extrapolation and generally 
has a lower precision as a result.
Aperture corrections were applied to the
\mg indices using
equation~2 of \citet{jfk95}.
The colours were corrected for reddening by deriving the
optical depth at $1 \, \rm \mu m$
from the $B-V$ colour excesses provided by \citet{sfd98}
(the precise method is described in \S~\ref{s_exmaf1}).
K-corrections were derived from the same
SED used to evaluate the reddenings.
As shown in Figure~\ref{colourmg},
the data reveal a tight correlation between the corrected
effective $V-I$ colour, designated by $(V - I)_e^0$, and the 
corrected \mg index.  
A linear least-squares fit gives
\begin{equation}
\label{eqcolmg}
(V-I)_e^0 = (1.05 \pm 0.10) \, \hbox{\mg} + (0.85 \pm 0.03)
\end{equation}
The rms scatter is only $0.04 \, \rm mag$.
For the same galaxies,
the relation between the total $V-I$ colour and the \mg index 
displays a much higher rms scatter of 0.09 mag.
In comparison, the effective $B-V$ colour, $(B-V)_e^0$,
varies with \mg as
\begin{equation}
\label{eqbvmg}
(B-V)_e^0 = (0.93 \pm 0.06) \, \hbox{\mg} + (0.64 \pm 0.02)
\end{equation}
Within errors, the slope is consistent with that found for
the relation between $(V-I)_e^0$ and \mg.
Thus, for a given uncertainty in \mg,
$(V-I)_e^0$ can be judged just as well as $(B-V)_e^0$.
The rms is less, though, being only 0.02 mag.
It is possible that the $V-I$ measurements employed
in determining $(V-I)_e^0$
are inherently more uncertain than the data which went
into determining $(B-V)_e^0$
because of the difficulties associated
with doing photoelectric photometry in $I$.

\subsection{Reddening Correction for the \mg Index}
\label{s_mgmaf1}

As mentioned in \S~\ref{mg2},
a measurement of the \mg index from the
spectrum of Maffei~1 is not directly comparable
with the \mg indices used to define the $(V-I)$--\mg relation,
as the reference \mg indices were
determined from spectra which are reddened to a much
lesser degree than that of Maffei~1.
To bring each \mg index for Maffei~1 on to the same
scale as the comparison \mg indices,
we obtained a first approximation of the
reddening of Maffei~1 using equation~\ref{eqcolmg}
with the \mg index measured from the
reddened spectrum. Next, we used this first
approximation to produce an unextinguished spectrum.
The unextinguished spectrum was then 
extinguished using the average reddening of the comparison
ellipticals.
The \mg index measured from this spectrum is therefore on
the same scale as the \mg indices used to define the
$(V-I)$--\mg relation. The revised value was inserted
into equation~\ref{eqcolmg},
from which we obtained a second approximation of the reddening
of Maffei~1. This process was repeated until
each \mg index converged.
Our final aperture-corrected values are \mgwxcor mag for the wide-slit spectrum
and \mgnxcor mag for the narrow-slit spectrum.
Thus, the reddening corrections lowered the \mg indices
by \mgxcor mag, which is almost twice the effect of the
aperture corrections.
We note that the corrected values are nearly identical to the
``true'' \mg indices,
i.e., the line strengths
we would measure if the galaxy were completely unobscured,
which are \mgwo mag and \mgno mag for the wide- and narrow-slit
spectrum, respectively.
This indicates that the extinction of the comparison
ellipticals is small enough that the shape of their
continua in the proximity of the \mg feature is not
noticeably altered by reddening.

To determine the intrinsic $V-I$ colour of Maffei~1
from equation~\ref{eqcolmg}, we adopt the \mg 
index measured from the narrow-slit spectrum,
namely \mgadopt mag, as opposed to the value
obtained from the wide-slit spectrum.
The reason for this is that the spectral
resolution of the narrow-slit spectrum (\resn) is 
comparable to the spectral resolution of the Lick-IDS system
($\rm 9 \, \AA$)
upon which the \mg indices used to construct the color--\mg
relation are based.
We note that the spectral resolution of the wide-slit spectrum 
is over twice
as poor as that of the narrow-slit spectrum, yet the \mg indices
obtained from the two spectra are in close agreement, differing
by only \mgdif mag. This indicates that the measurement of the \mg index
is not strongly dependent on spectral resolution within the
range of resolutions in our observations.

\subsection{The Extinction of Maffei~1}
\label{s_exmaf1}

From equation~\ref{eqcolmg}, the fully-corrected \mg index obtained
from the narrow-slit spectrum of Maffei~1 
implies $(V-I)_e^0 = \,$ \colVIe mag.
Adopting $(V-I)_e = 3.12 \pm 0.05 \rm \, mag$ 
\citep{bm99},
we obtain $E(V-I)=$ \EVI mag.
As well, equation~\ref{eqbvmg} gives $(B-V)_e^0=$ \colBVe mag.
Since $(B-V)_e = 2.42 \pm 0.08$ \citep{bm99},
then $E(B-V)=$ \EBV mag.

The measured colour excess can be used to judge the extinction
in any passband provided that appropriate account is taken for
the shifts in effective wavelength as a function of source
morphology, reddening, and redshift. \citet{ma00} and 
\citet{mlm02} have developed procedures for doing so.
First, a monochromatic extinction curve appropriate for the
diffuse interstellar medium, which is primarily responsible
for obscuring extragalactic sources, must be adopted.  
\citet{fi99} has developed an algorithm for determining the
monochromatic reddening law associated with any particular value
of the ratio of total to selective extinction, $R_V = A_V / E(B-V)$.
For the diffuse interstellar medium, $R_V = 3.07 \pm 0.05$
for a star of zero colour in the limit of zero extinction
\citep{ma00}.  The reddening law chosen to evaluate the
extinction of Maffei~1 is that given by the algorithm of 
Fitzpatrick which, when applied to the spectrum of Vega,
yields $R_V = 3.07$ after integrating the flux passed by
response curves characterizing $B$ and $V$ filters.

The extinction of a source is best quantified by its optical
depth at some wavelength.  Knowing optical depth, it is
possible to evaluate the extinction of a source in any passband by
applying the appropriately scaled reddening law to the spectrum
and integrating through the response function for the filter.
The best choice of wavelength is $1 \, \rm \mu m$, as it is
in a part of the reddening law which is not very sensitive to
environment, and because the optical depth there is comparable
numerically to $E(B-V)$.  The optical depth at $1 \, \rm \mu m$,
$\tau_1$, can be computed from any colour excess by an iterative process.
First, a value for $\tau_1$ is guessed, and the spectrum of the 
source is extinguished with the aid of the scaled reddening law.  
For each broad-band filter $\Lambda$ used to compute the colour 
excess, the ratio $R^1_\Lambda$ of the total extinction $A_\Lambda$ 
to the optical depth $\tau_1$
is determined by integrating the spectrum
transmitted by the response curve for the filter.  Then,
the value of $\tau_1$ is revised, and the process repeated.
Convergence is generally rapid.

To determine $\tau_1$ for Maffei~1, \citet{mlm02} created
a truly integrated SED
for a typical unreddened
elliptical galaxy extending all the way from 
$0.13$ to $3.8 \, \rm \mu m$.
This was accomplished by combining spectra and photometry 
available for NGC~3379.  The spectrum was shifted to the
redshift of Maffei~1, and then iteratively reddened until
the observed value of the colour excess for Maffei~1 was 
reproduced.  Filter response curves were adopted from
\citet{bes90}.  In the case of $E(V-I)$, the analysis
converged on a value of
$\tau_1$ equal to $\opd \pm \dopd$, which leads to
$A_V = $ \AV mag.  Note that the error here does
not include that in the zero-point of the reddening law
(i.e., $R_V$ for Vega).
In comparison, $E(B-V)$ yields an optical depth of \opdEBV,
from which $A_V = $ \AVEBV mag.
These values are consistent with the results from $E(V-I)$,
justifying the choice of reddening law.  However, 
because of the added uncertainty
in the photometry in $B$, the errors are twice as great.
For this reason, along with the fact that the Galactic
reddening law is more sensitive to environment in the ultraviolet,
we choose to adopt for $\tau_1$
the value derived from $E(V-I)$ alone.
The corresponding
values of $R_\Lambda^1$  and $A_\lambda$
for $B$, $V$, and $I$ (Cousins)
are given in Table~\ref{tabmaf}.

If instead the non-monochromatic
reddening law of \citet{ccm89} is employed
(parameterized as before to deliver $A_V / E(B-V)$ equal to 3.07
for Vega),
then $E(V-I)$ and $E(B-V)$ lead to values of $\tau_1$ equal
to $\rm \opdEVIccm \, and \, \opdEBVccm$, respectively.
The two values are significantly
more discrepant than those derived using the law of \citet{fi99}.
The implied difference in $A_V$ amounts to \AVdifccm mag,
as against \AVdiffitz mag based upon the analysis with
the Fitzpatrick law.

Our adopted
extinction for Maffei~1 in $V$ is \AV mag.
This is consistent with the estimate
of \citet{dav02} which,
after corrections for effective wavelength shifts
(in the manner described for Maffei~1,
but with the aid of the spectrum of a $\rm M0~III$ star)
leads to $A_V=$ \AVdav mag.
Our result is also consistent with that required by
the column density of gas
along the line of sight, which is $A_V \rm = 4.5 \pm 0.4 \, mag$
after correcting for effective
wavelength shifts using the spectrum of a $\rm B0~V$ star. 
It is \AVdifbmcol mag less than demanded by the $B-V$
colour comparison of \citet{bm83}
which, after correcting for effective wavelength shifts,
is $A_V \rm = 4.9 \pm 0.2 \, mag$.

\section{Distance}
\label{distance}

Until the late 1980s,
distances to elliptical galaxies were 
judged from the Faber-Jackson relation,
which correlated an elliptical galaxy's 
luminosity with its velocity dispersion.
Unfortunately, the scatter in the Faber-Jackson 
relation was at least twice that of the 
Tully-Fisher relation for spirals,
on the order of 0.8 mag in $M_B$.
However, independent studies by \citet{dd87} 
and \citet{dlb87} revealed that the addition of the 
surface brightness as a third parameter considerably 
tightened the relation.
This led to the realization that elliptical galaxies are 
confined to a plane, called the ``Fundamental Plane (FP)'',
in a three-dimensional 
space defined by coordinates linked to
mass, density, and the mass-to-light ratio.
One way of describing the Fundamental Plane is through
the relationship among
the metric effective radius $R_e$,
the velocity dispersion $\sigma$,
and the mean surface brightness $<{\cal I}>_e$ within $R_e$:
\begin{equation}
\label{eqfp}
R_e \propto \sigma^\alpha <{\cal I}>_e^{-\beta}
\end{equation}
Independent determinations of the power-law 
exponents for different passbands have found
$\alpha = 1.2$--1.4 and $\beta = 0.7$--0.9, 
with no obvious trend of the coefficients with wavelength.
This form of the Fundamental Plane is adopted to 
evaluate the distance to Maffei~1.

Because of the reduced extinction and the
availability of wide-field surface photometry,
the appropriate passband to use to derive an FP distance to
Maffei~1 is $I$.  Unfortunately, the FP
has never been constructed in this passband.  We do so here
using spectroscopy and $I$-band photometry
obtained by \citet{sgha,sghb}
for a sample of 17 elliptical galaxies in the Coma Cluster.
The Coma cluster is ideal for a FP analysis,
as its large distance minimizes scatter due to the
varying depths of galaxies within the cluster.
Various independent studies validate 
the assumption that the FP in 
this distant realm has the same size and shape 
as that for nearby clusters \citep{jfk96,kel00}.
Furthermore,
\citet{rdz01} find no dependency of the
FP on environment,
based on a comparison of ellipticals in compact groups
with a sample of ellipticals in the field or in
very loose groups.

For the distance to Coma, we adopt
$78.8 \, \rm Mpc$.  This is the average value measured from 
the FP in Gunn $r$ and from the Tully-Fisher relation
in $I$ (which are consistent within 0.2 Mpc; see \citealt{fmg01}),
but anchored to the maser distance of NGC~4258 (7.2~Mpc:
\citealt{hmg99}) rather than
to the Cepheid distance of the LMC \citep[see][]{fmg01}.

The elliptical sample is given in Table~\ref{tabfp}.
Great care was taken to correct observables in the same manner
as Maffei~1.
After first transforming effective
radii to effective aperture radii following the prescription
of \citet{ov81},
velocity dispersions were corrected to the
standard aperture
using equation~2 of \citet{jfk95}, 
which is appropriate for 
$\rm -1.4 \leq log(\theta_{obs}/\theta_e) \leq -0.5$.
Extinctions were estimated
by extracting $E(B-V)$ from the maps
of \citet{sfd98}, and then deriving $\tau_1$ iteratively using
the SED of an elliptical galaxy.
Surface brightnesses were corrected for the K-effect
and $(1 + z)^4$ dimming, the former determined from the same
SED used to arrive at $\tau_1$.  
In an effort to lessen scatter due
to varying orientations, 
surface brightnesses were also multiplied by the axis
ratio $q$ derived from $\log R_{25}$ in $B$ \citep{vau91}. 
For true oblate spheroids, this
reduces surface brightnesses 
to the values that would be seen if the galaxies 
were observed face-on.

The $I$-band FP for the  Coma
sample is plotted in Figure~\ref{figfp}.
Note that $R_e$ is the metric length in kpc of the semi-major axis
of the elliptical isophote encompassing half of the total
light in $I$ (the effective isophote),
$\sigma_{e8}$ is the aperture-corrected velocity dispersion 
in $\kms$
(see \S~\ref{apcor}),
and $<\mu>_e^0$ is the fully-corrected estimate for the mean
face-on surface brightness within $R_e$, in $\rm mag \, arcsec^{-2}$
(see Table~\ref{tabfp}).
Values of $<\mu>_e^0$ were derived
from the surface brightness of the effective isophote,
$\mu_e$, and values of $R_e$ were calculated from the
length of the semi-major axis of that isophote, $r_e$.  
Both $r_e$ and $\mu_e$ were measured by fitting free
ellipses to the surface brightness profiles.
A linear least-squares fit to the data gives
\begin{equation}
\label{eqfpcoma}
\log R_e = (0.87 \pm 0.19) \log \sigma_{e8} + 
(0.83 \pm 0.06) <\mu>_e^0 / 2.5 - (7.79 \pm 0.69)
\end{equation}
The rms scatter in the
relation is $0.09$ in $\log R_e$, as against $0.11$
when the axis ratio correction is excluded.
The fit to the data without the axis
ratio correction yields 
$\alpha=1.18 \pm 0.24$ and $\beta=0.90 \pm 0.08$.  These values
are consistent within errors with 
the results of \citet{jfk96}, who,
from photometry in Gunn $r$,
found $\alpha=1.24 \pm 0.07$ and $\beta=0.82 \pm 0.02$.

The FP parameters for Maffei~1 are given at the bottom of
Table~\ref{tabfp}.
The effective radius $r_e$ and effective 
surface brightness $\mu(r_e)$ were
measured by \citet{bm99} via a free-ellipse fit to 
the $I$-band surface brightness profile,
exactly as was done for the ellipticals 
used to construct the FP.
The surface brightness was corrected for inclination 
using the mean axis ratio of the 
outer isophotes in $I$
\citep{bm99}.
This quantity was used in place 
of $\log R_{25}$ to avoid the uncertainty introduced by the 
heavy extinction in the $B$-band.
Based upon equation~\ref{eqfpcoma},
$\log R_e = \,$ \logRe for Maffei~1.  Thus,
the FP distance to Maffei~1 
is \FPD Mpc.
The error comes from the uncertainty
in the fit and the errors in the measurements of $r_e$, $\mu_e$,
and $\sigma$ for Maffei~1.  The uncertainty in the 
distance to the Coma Cluster is not included.

Since there is now wide-field surface photometry
available in $B$ \citep{bm99}, it is possible to determine the distance
to Maffei~1 with some confidence from the $D_n-\sigma$ relation, too
\citep{lfb88}.
It is of benefit to do so because the photometry 
involved and the calibration of the relation are completely
independent of those employed in our FP analysis in $I$.
The technique relates the Hubble velocity $v_H$
to the velocity dispersion $\sigma$ and the 
diameter $D_n$ within which the mean $B$-band 
surface brightness is $\rm 20.75 \, mag \, arcsec^{-2}$.
From \citet{lfb88},
\begin{equation}
\label{eqdnsigma}
\log v_H = 1.2 \log \sigma - \log D_n + 1.411
\end{equation}
where $v_H$ is in $\kms$,
$\sigma$ is in $\kms$,
and $D_n$ is in $0\farcm1$.
For $\sigma$, we have \VdapDav $\kms$,
which is
the velocity dispersion measured in \S~\ref{vdisp} after 
the aperture correction of \citet{dsp93},
instead of \citet{jfk95}, for consistency with
the definition of this parameter
by \citet{lfb88}.
We note that this value is
$\rm 24 \, \kms$ less than the value of $\rm 200 \pm 50 \, \kms$
adopted by \citet{lt93} in their $D_n-\sigma$ analysis.
For the $B$-band extinction, 
we use $A_B=$ \AB mag,
derived from the value of $\tau_1$ obtained here
using $R_B^1 = A_B / \tau_1 = \RB$.
The surface photometry of \citet{bm99} gives
$D_n = $ \DnB.
Application of equation~\ref{eqdnsigma} 
yields $v_H = $ \DnBv $\kms$,
which we increase by $17 \%$ to compensate for
``Malmquist bias'' according to the prescription
of \citet[see equation~2.11]{lfb88}.
The Hubble constant is
$79.2 \pm 9.9 \, \kms \, Mpc^{-1}$ based upon
the Cepheid calibration of \citet{fmg01},
re-anchored to the maser distance of NGC~4258
\citep[see][]{nfs01}.
Thus, the corrected velocity corresponds to a distance of
\DnDB Mpc, where once again the uncertainty
in the zero-point of the distance scale has been excluded.
The large error bar
is predominantly due to the uncertainty in 
the $B$-band photometry and extinction.

A value of $D_n$ can be estimated as well from the
$K^\prime$ photometry of \citet{lt93}, but using an actual
measurement of $B - K^\prime$ now.
The effective $B$ magnitude is given by \citet{bm99}, and
the effective $K^\prime$ magnitude within the effective
aperture defined for $B$ \citep{bm99} can be measured from
the aperture growth curve of \citet{lt93}.
The resulting $B-K^\prime$ colour is \colBKt mag, 
versus $4.2 \, \rm mag$ adopted by \citet{lt93}.
From the $K^\prime$ surface brightness profile of \citet{lt93},
we measure $D_n = $ \DnK using the revised colour.
Performing the same analysis as above,
we obtain a $D_n-\sigma$ distance of \DnDK Mpc.

Although distances determined from the brightest stars in 
$K^\prime$ are suspect, we have no grounds to reject the
new values from the $D_n-\sigma$ relation.  Consequently,
the distance we adopt for Maffei~1 is \aveD Mpc, 
which is the weighted average
of the result from the FP and the mean
of the two estimates from the $D_n-\sigma$
relation.  The error does not include that in the zero-point
of the distance scale.
It also does not include the uncertainty in the shape of
the reddening law, but this is thought to be small
in comparison to the error in the measurement of
the extinction (see \citealt{ma00} and \citealt{bm02}).

\section{Discussion}
\label{discussion}

To properly compare our result with
the distance estimates of \citet{lt93}
and \citet{db01} derived from the brightest stars in $K^\prime$,
we revise their results
using the Galactic extinction of Maffei~1 found in this study 
and our adopted distance scale anchored to the maser distance
of NGC~4258. 
The result of \citet{lt93} moves in
from $4.2 \pm 1.1 \, \rm Mpc$ to \SBFDlt Mpc, and the
result of \citet{db01}
comes in from $4.4^{+0.6}_{-0.5} \rm \, Mpc$ to \SBFDdb Mpc.
Our new distance is significantly
closer.

As stated in \S~\ref{intro},
both of the $K^\prime$ analyses required the adoption of 
an intrinsic luminosity for the brightest giants in $K^\prime$,
which was judged from observations of M32 and the bulge of M31.
Luminosities of stars at the tip of the AGB
are very sensitive to age 
\citep{bre94};
over 2--14 Gyr, $M_{K^\prime}$ becomes fainter by roughly $1 \, \rm mag$.
M32 is known to host a population of luminous 
AGB stars, evidence of an episode of 
star formation in the last 2 Gyr 
\citep{fre92,bre94}.  The bulge of M31 may be similarly affected,
given that the very brightest stars at $K^\prime$ have
absolute magnitudes close to those in M32 \citep{dav01}.
In Maffei~1, the lack of an unusually large colour gradient 
\citep{bm99} and the negligible \ha emission \citep{bm02}
suggest that star formation has not occurred recently.
Consequently, it
may be postulated that AGB stars in Maffei~1 
are more likely to lie on the opposite 
end of the age scale from their counterparts in M31 and M32.
If the age of the brightest AGB stars in Maffei~1 were around 10 Gyr,
whereas the age of such stars in M32 and in the bulge of M31 were more
like 2 to 4 Gyr,
absolute $K^\prime$ magnitudes
would be fainter in Maffei~1 by roughly $0.5 \, \rm mag$.
With the bulge of M31 as the reference field,
the distance of \citet{db01} would be lowered to \SBFDagedb Mpc, and
the \citet{lt93} distance would be reduced 
to \SBFDagelt Mpc. Both of these values are in line with
our measurement.

The role that the three dominant members of the IC~342/Maffei group
have played in the evolution of the Milky Way 
and its neighbours has been 
a subject of controversy due to their uncertain distances.
Previous estimates of the radial 
velocities and distances to Maffei~1 and IC~342 suggested 
that these galaxies could have been 
in the vicinity of the Local Group within the last 5 to 8 Gyr
\citep{mlm89, vbm93}, in the process
affecting its early dynamical evolution.
This conclusion is all the more significant given that a
fundamental tenet of Local Group timing is that 
the Milky Way and M31 have always behaved dynamically
like an isolated pair.

The new velocity and distance for Maffei~1 make possible a
new judgment about the galaxy's relevance to the Local Group.
With a heliocentric velocity of $\Vhel \, \kms$,
the velocity
of Maffei~1 with respect to the centroid of the Local Group
is $330 \, \kms$ \citep{cvb99}.
Adopting $H_0=79.2 \, \kms \, Mpc^{-1}$,
the Hubble distance to
Maffei~1 is $4.1 \, \rm Mpc$.
This is somewhat higher than the distance computed above,
but it is easily possible that it is distorted by the peculiar
motion of Maffei~1 arising from its association
with the IC~342/Maffei Group. 
The mean velocity
of the three dominant galaxies relative to the Local Group 
is $\vGrp \, \kms$,
where the recessional velocities of Maffei~2 and IC~342
have been adopted to be those given by
\citet{hth96} and \citet{new80}, respectively.
Using this velocity,
the Hubble distance comes
down to \HubD Mpc, which is marginally consistent with our
result.  We therefore conclude that
it is unlikely that Maffei~1 interacted
with the Local 
Group since the Big Bang.
Further discussion of this issue will be given in \citet{fml03},
where modern distances to IC~342 and Maffei~2 will be derived
and shown to be close to that derived here for Maffei~1.

\section{Summary}
\label{summary}

New spectroscopy of Maffei~1 at low 
and high resolution has been presented.
The \mg index has been measured for the first time, 
and the heliocentric velocity and
velocity dispersion have been derived with improved accuracy.

The \mg index was used to re-evaluate the extinction.
To avoid the uncertainties inherent in $B$-band
photometry of Maffei~1, the relationship between \mg 
and $V-I$ for elliptical galaxies was examined
by constructing a sample of relatively unobscured objects with
existing photometry.  A strong correlation was discovered,
which was employed to estimate the intrinsic $V-I$ colour, and, thus,
$E(V-I)$ for Maffei~1.
Taking care to compensate for the effect of extinction
on the \mg index and on shifts
in effective wavelengths,
the optical depth at $1 \, \rm \mu m$ due to galactic
dust was estimated to be
$\tau_1 = \opd \pm \dopd$.  Integrating an appropriately extinguished
SED for an elliptical galaxy through the
response curve for the $V$ filter,
$A_V=$ \AV mag.
The value of $A_V=5.1 \pm 0.2 \, \rm mag$ found by \citet{bm83}
was an overestimate because of inadequate or incorrect compensation
for effective wavelength shifts.

The distance of Maffei~1 was derived by constructing the
Fundamental Plane for the Coma Cluster in $I$,
then taking the weighted average of the Fundamental Plane distance
and the mean of two estimates from the $D_n$--$\sigma$
relation.
The result is \aveD Mpc. It is
lower than the estimates of \citet{lt93} and \citet{db01},
which are founded upon the brightest stars
seen at $K^\prime$.
We suggest that this discrepancy
is due to their assumption of similar stellar 
populations among Maffei~1, M31 and M32,
despite indications to the contrary.

The revised properties of Maffei~1 
are summarized in Table~\ref{tabmaf}.
The apparent magnitude in $V$ after correction 
for extinction shows that Maffei~1 
is among the brightest galaxies in the Northern sky. 
Knowing the extinction, the apparent $B-I$ colour
and the surface brightness profile in $I$ can be used to
judge the unextinguished diameter at the standard isophote 
(i.e., $25 \, \rm mag \, arcsec^{-2}$ in $B$).
Based upon the fixed-ellipse profile of \citet{bm99}, 
the diameter of Maffei~1 is \Do, which is
two-thirds that of the full Moon.

The only other nearby galaxy in the class of Maffei~1 is Centaurus~A.
\citet{mar00} give a detailed summary of distance estimates to
this galaxy and adopt the weighted mean. We modify their
result by shifting the distance to the NGC~4258
zero-point adopted throughout this paper
\citep{fmg01,nfs01}.
We also update the estimate
obtained from surface
brightness fluctuations (SBF) in $I$ \citep{isr98}, which are known 
to be a more robust distance indicator than fluctuations in $K^\prime$.
\citet{ton01} present a new $I$-band calibration, which
we apply to the fluctuation magnitude measured by \citet{ton90}.
The resulting distance modulus is \dmodCenA mag
corrected to the
NGC~4258 zero-point.
In deriving this value,
the extinction corrections employed
were obtained from $E(B-V)$ \citep{sfd98} 
in the manner described for Maffei~1.
After substituting this result for the previous SBF distance
adopted by \citet{mar00},
we obtain a weighted mean distance modulus of
$27.77 \pm 0.07 \rm \, mag$, which corresponds to
$3.6 \pm 0.1 \rm \, Mpc$.
The distance estimates to Maffei~1 from the Fundamental Plane
and $D_n$--$\sigma$ relation suggest that Maffei~1 is closer,  making it
the nearest giant elliptical galaxy to the Milky Way.
However, based upon
the total apparent magnitude
given in the Lyon Extragalactic
Database ($V_T = 6.72$), the absolute magnitude of Cen~A is
$M_V = \MVCenA \, \rm mag$. Maffei~1 is fainter by
$\MVdif \, \rm mag$.

\acknowledgments

MLM gratefully acknowledges the
continuing support of the Natural Sciences and Engineering Research
Council of Canada.  He is also grateful to R. Ross for performing the initial data reduction, and to J. Kormendy for his assistance
during the early stages of this work.
RB acknowledges the support of NASA/STScI grant
GO 8101 to the University of Alabama.
The authors would like to thank the referee for
very constructive comments, which have helped to
strengthen the methods behind the results of this paper.

\clearpage

\figcaption[]{
\label{fighispec}
High-resolution spectra of Maffei~1 (bottom) and 
the K2 giant HD~22427 (top),
one of four template stars used to measure 
the radial velocity and velocity dispersion.
The \mg absorption feature 
can be clearly seen at $\rm \lambda\lambda5155-5198 \, \AA$.
}

\figcaption[]{
\label{figlospec}
Observed (a) and reddening-corrected (b)
narrow-slit spectra of Maffei~1.
The central region marks the 
boundaries of the \mg absorption feature.
The continuum level is defined by
the mean flux within the blue and red bands 
marked by the outer shaded regions.
The reddening-corrected spectrum illustrates
what would be observed if Maffei~1
were reddened on average as much as
the comparison ellipticals
(see \S~\ref{s_mgmaf1}).
}

\figcaption[]{
\label{colourmg}
Relationship between effective $V-I$ 
colour and \mg index for elliptical galaxies.
The solid line is the linear least-squares fit. 
The rms scatter in $\rm (V-I)^0_e$ is 0.04 mag.
The dashed line marks the \mg 
index of Maffei~1.}

\figcaption[]{
\label{figfp}
$I$-band Fundamental Plane. The solid line is the 
linear least-squares fit. The rms 
scatter in $\log R_e$ is 0.09.
The dashed line marks the location of Maffei~1.}

\newpage
\plotone{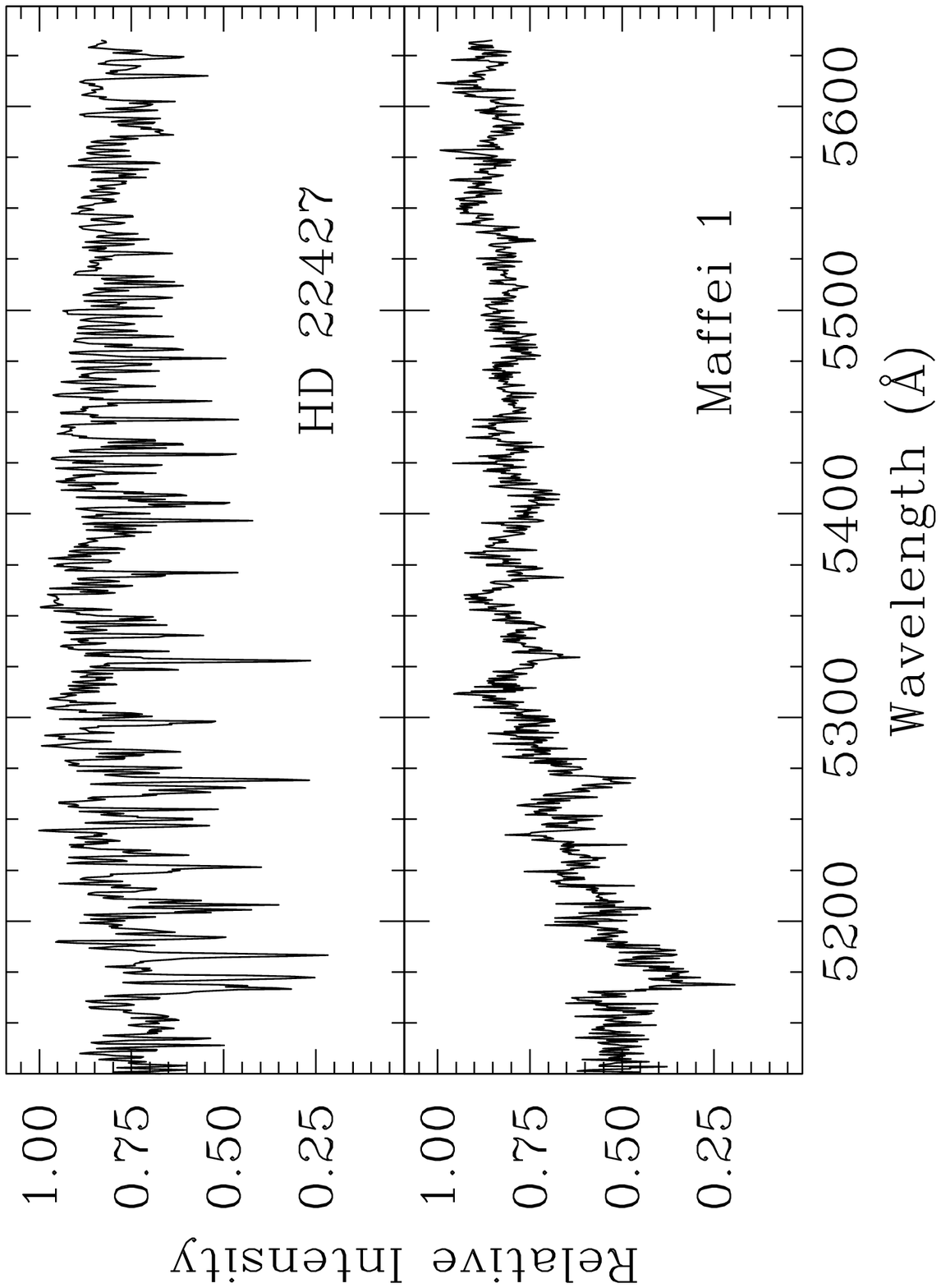}
\newpage
\plotone{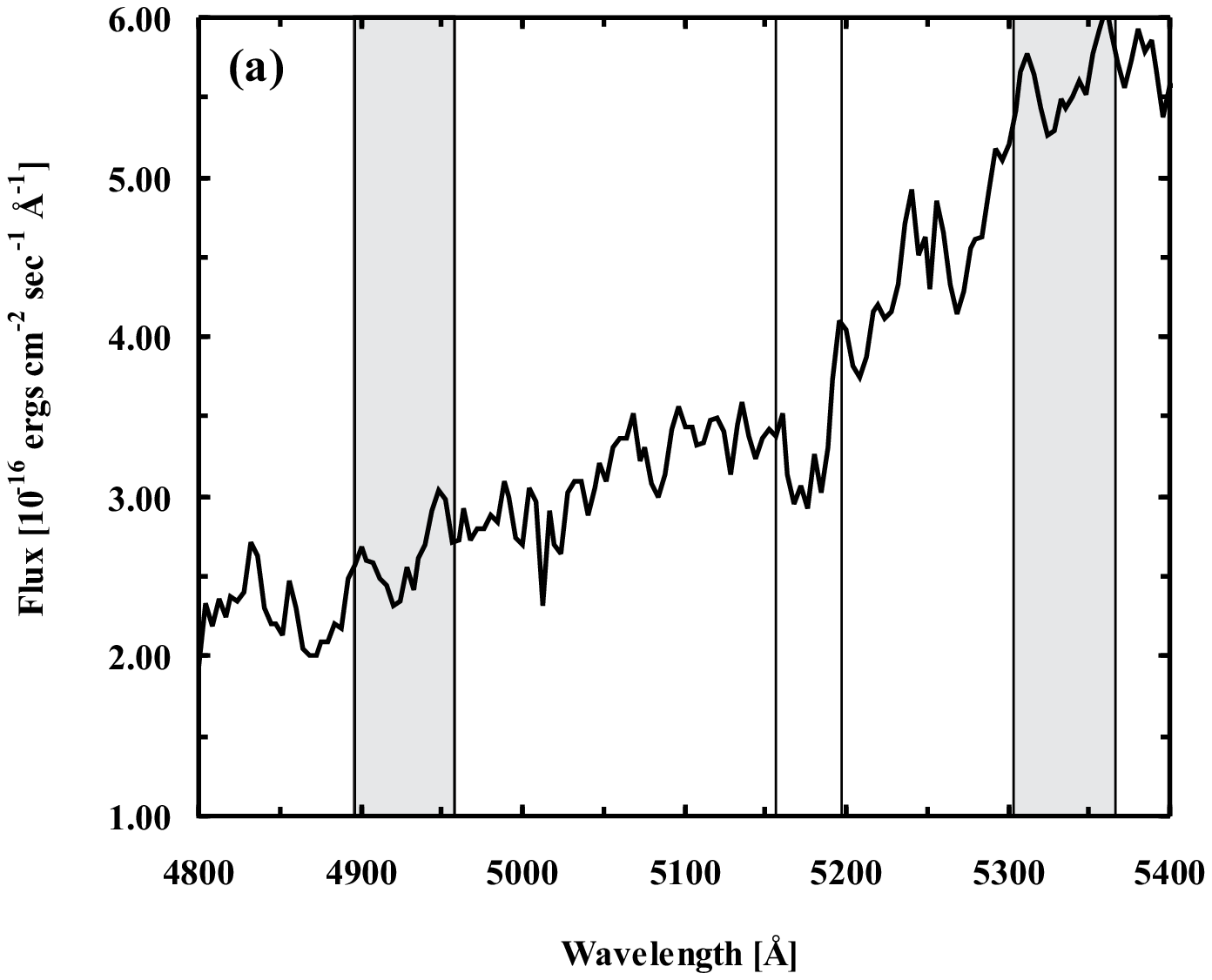}
\newpage
\plotone{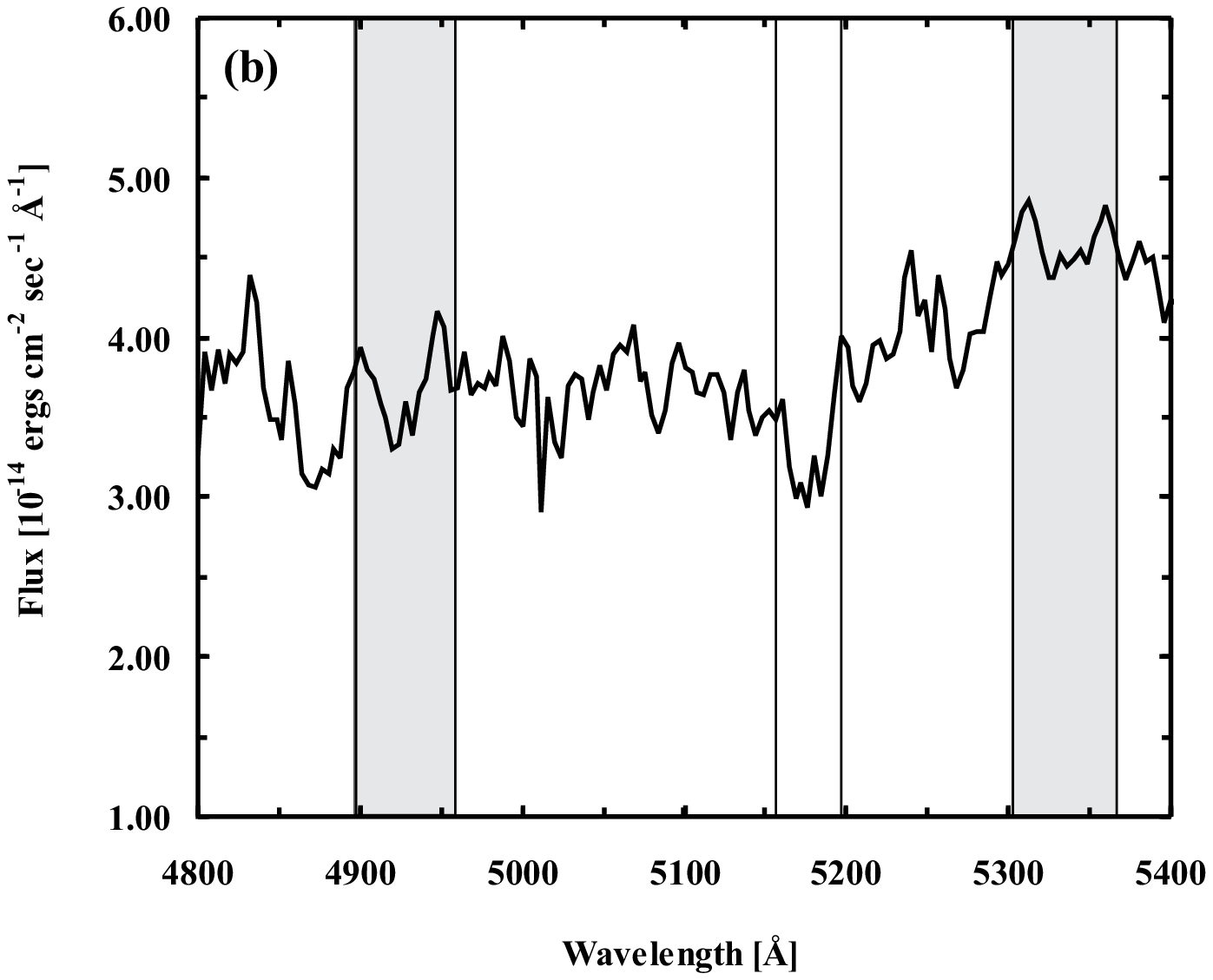}
\newpage
\plotone{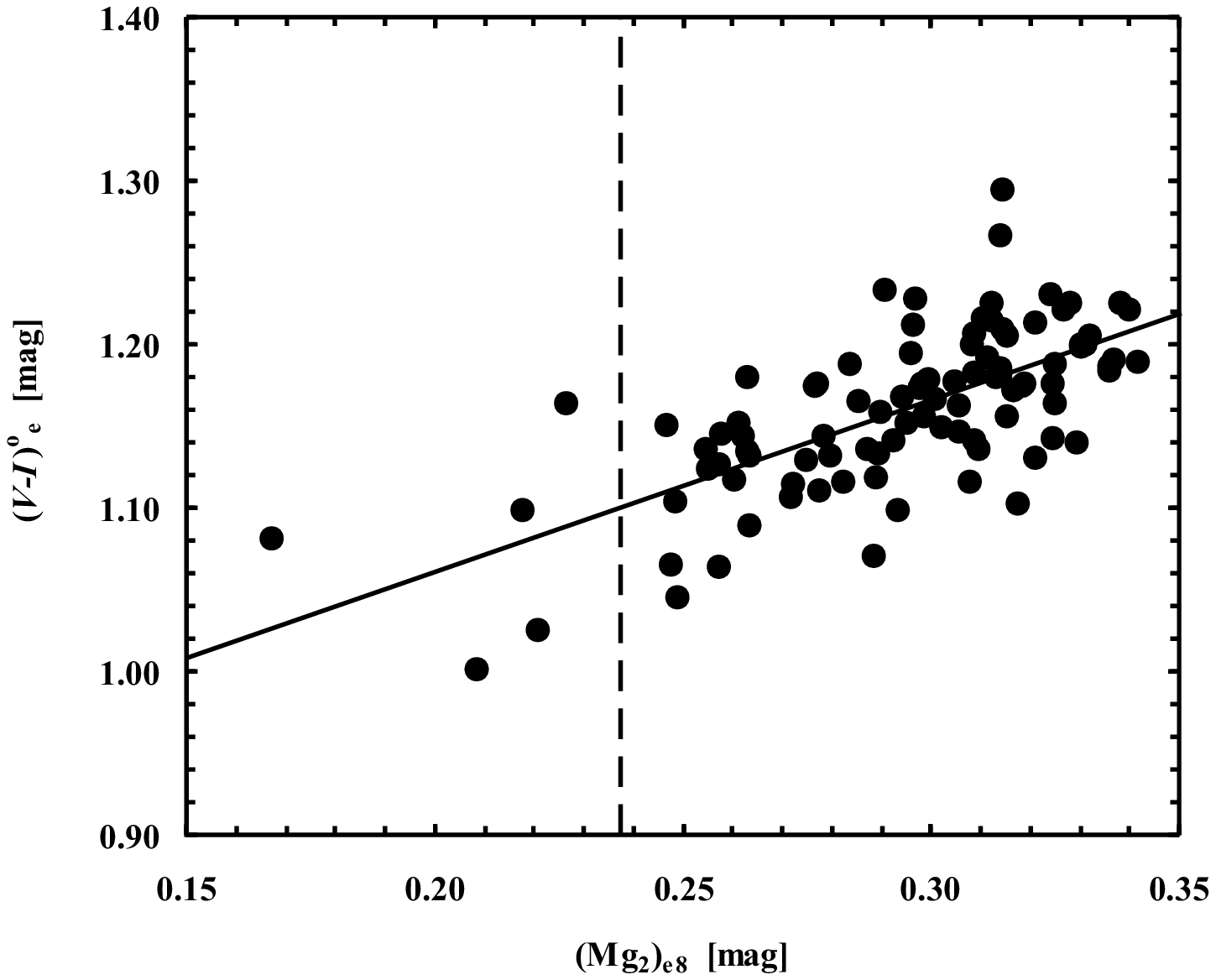}
\newpage
\plotone{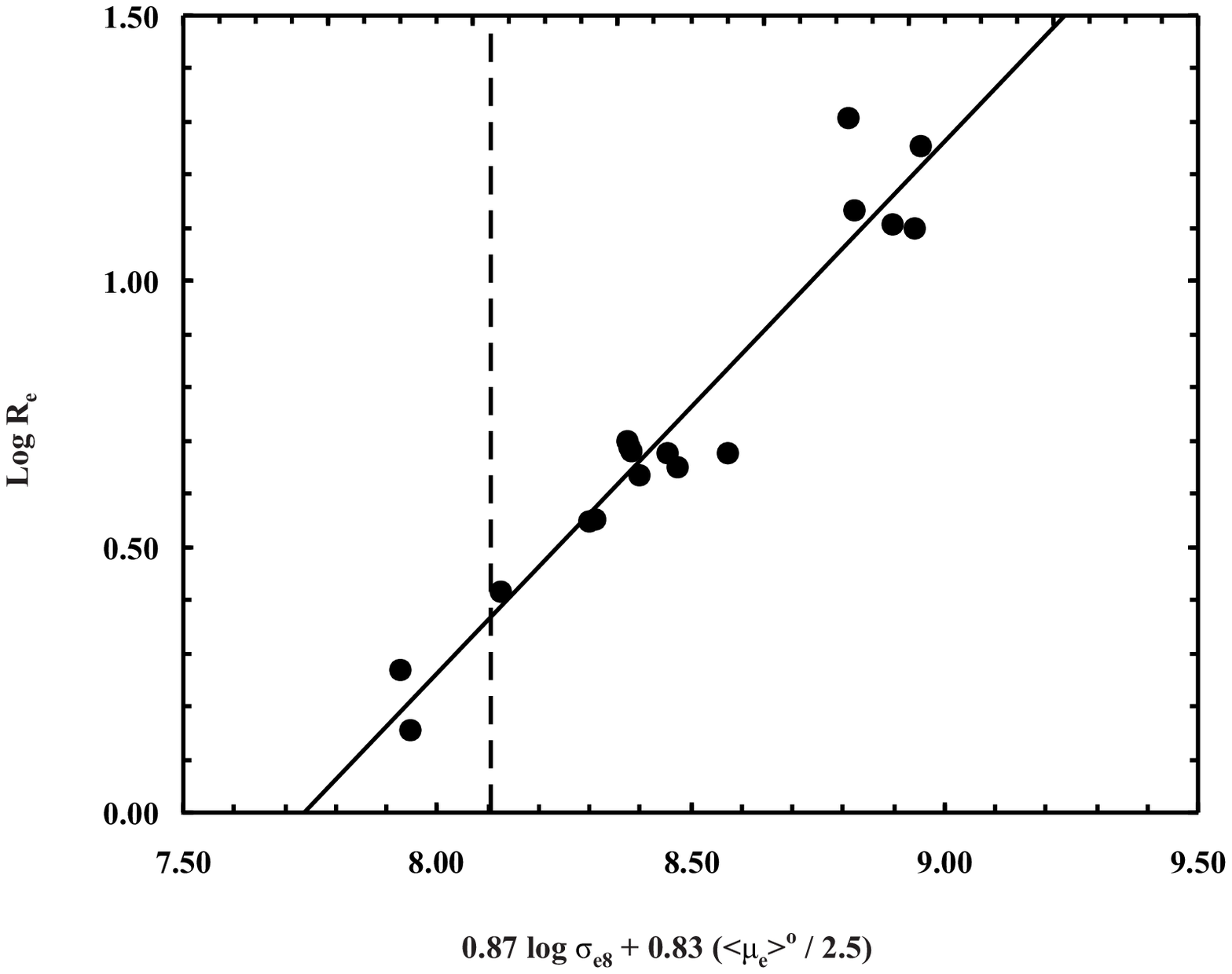}

\clearpage
\begin{deluxetable}{lccccccc}
\tablewidth{0pt}
\tablecaption{
\label{tabkin}
Velocity Measurements
}
\tablehead{
\colhead{Template} & 
\colhead{Type} &
\colhead{$v_\odot$} & 
\colhead{Refs} &
\colhead{$v_\odot$ (Maf 1)} & 
\colhead{Error} &
\colhead{$\sigma_r$ (Maf 1)} &
\colhead{Error}
\\
\colhead{} &
\colhead{} &
\colhead{($\rm km \, s^{-1}$)} &
\colhead{} &
\colhead{($\rm km \, s^{-1}$)} &
\colhead{($\rm km \, s^{-1}$)} &
\colhead{($\rm km \, s^{-1}$)} &
\colhead{($\rm km \, s^{-1}$)}
\\
\colhead{(1)} &
\colhead{(2)} &
\colhead{(3)} &
\colhead{(4)} &
\colhead{(5)} &
\colhead{(6)} &
\colhead{(7)} &
\colhead{(8)}
}

\startdata

HD 5459, \#1 & G8~IV     & \phn $-$9.4  & 1,4 & 63.7 & 14.5 & 184.5 & 17.4 \\
HD 9166      & K3~III    & $-$13.9      & 2,5 & 71.1 & 12.8 & 191.9 & 15.4 \\
HD 5459, \#2 & G8~IV     & \phn $-$9.4  & 1,4 & 70.3 & 14.8 & 183.7 & 17.7 \\
HD 22427     & K2~III-IV & $-$31.5      & 3,6 & 60.7 & 13.3 & 189.4 & 15.9 \\
HD 27224     & K1~III    & \phn $-$8.3  & 3,7 & 85.5 & 14.2 & 184.3 & 17.1 \\
\cline{1-8} \\
Adopted & \nodata & \nodata & \nodata & \Vhel & \Vhelerr & \Vd & \Vderr

\tablecomments{
(1) Name of template star, in order of time of observation.
(2) Spectral type of template star.
(3) Heliocentric radial velocity adopted for template star.
(4) References for spectral type and radial velocity.
(5) Heliocentric radial velocity derived for Maffei 1.
(6) Uncertainty in velocity of Maffei 1 (see \S~\ref{vdisp}).
(7) Velocity dispersion derived for Maffei 1.
(8) Uncertainty in velocity dispersion of Maffei 1 (see \S~\ref{vdisp}).
}

\tablerefs{
(1) \citealt{feh61};
(2) \citealt{rom55};
(3) \citealt{sch71};
(4) \citealt{eva67};
(5) \citealt{wil53};
(6) \citealt{gri70};
(7) \citealt{feh87}
}

\enddata
\end{deluxetable}

\clearpage
\begin{deluxetable}{lcccccccccc}
\tabletypesize{\scriptsize}
\tablewidth{0pt}
\tablecaption{Fundamental Plane Parameters\label{tabfp}}

\tablehead{
\colhead{Galaxy} & 
\colhead{$v_\sun$} &
\colhead{$b/a$} &
\colhead{$r_e$} &
\colhead{$\rm \theta_e$} &
\colhead{$\rm <\mu>_e$} &
\colhead{$\tau_1$} &
\colhead{$A_I^{gal}$} &
\colhead{$A_I^c$} &
\colhead{$\rm <\mu>_e^0$} &
\colhead{$\rm \sigma_{e8}$}
\\
\colhead{} & 
\colhead{($\kms$)} &
\colhead{} &
\colhead{($\arcsec$)} &
\colhead{($\arcsec$)} &
\colhead{(mag $\rm arcsec^{-2}$)} &
\colhead{} &
\colhead{(mag)} &
\colhead{(mag)} &
\colhead{(mag $\rm arcsec^{-2}$)} &
\colhead{($\kms$)}
\\
\colhead{(1)} & 
\colhead{(2)} & 
\colhead{(3)} & 
\colhead{(4)} & 
\colhead{(5)} & 
\colhead{(6)} & 
\colhead{(7)} & 
\colhead{(8)} & 
\colhead{(9)} & 
\colhead{(10)} & 
\colhead{(11)} 
}

\startdata
IC 3957 & 6350 & 1.00 & 3.73 & 3.73 & 18.48 & 0.01 & 0.02 & 0.10 & 18.34 & 157 \\
IC 4041 & 7114 & 1.00 & 12.59 & 12.59 & 20.14 & 0.01 & 0.02 & 0.11 & 19.99 & 117 \\
NGC 4798 & 7838 & 0.76 & 46.96 & 41.27 & 21.20 & 0.01 & 0.02 & 0.13 & 21.34 & 163 \\
NGC 4807 & 6941 & 0.81 & 6.80 & 6.16 & 18.34 & 0.01 & 0.02 & 0.11 & 18.42 & 236 \\
NGC 4821 & 6974 & 0.62 & 11.69 & 9.43 & 19.44 & 0.01 & 0.02 & 0.11 & 19.82 & 173 \\
NGC 4827 & 7650 & 0.87 & 12.45 & 11.65 & 19.18 & 0.01 & 0.02 & 0.12 & 19.18 & 293 \\
NGC 4839 & 7346 & 0.49 & 33.07 & 24.57 & 20.23 & 0.01 & 0.02 & 0.12 & 20.85 & 241 \\
NGC 4842 & 7297 & 1.00 & 4.87 & 4.87 & 18.10 & 0.01 & 0.02 & 0.12 & 17.95 & 212 \\
NGC 4849 & 5885 & 0.78 & 13.04 & 11.58 & 19.12 & 0.01 & 0.02 & 0.09 & 19.26 & 218 \\
NGC 4859 & 7055 & 0.63 & 12.43 & 10.12 & 19.49 & 0.01 & 0.02 & 0.11 & 19.84 & 221 \\
NGC 4874 & 7175 & 1.00 & 52.91 & 52.91 & 20.86 & 0.01 & 0.02 & 0.11 & 20.71 & 195 \\
NGC 4886 & 6345 & 1.00 & 9.21 & 9.21 & 19.59 & 0.01 & 0.02 & 0.10 & 19.46 & 151 \\
NGC 4889 & 6497 & 0.65 & 35.74 & 29.37 & 19.61 & 0.01 & 0.02 & 0.10 & 19.95 & 392 \\
NGC 4919 & 7110 & 0.62 & 9.34 & 7.54 & 18.97 & 0.01 & 0.02 & 0.11 & 19.35 & 169 \\
NGC 4931 & 5849 & 0.39 & 11.32 & 7.84 & 18.46 & 0.01 & 0.02 & 0.09 & 19.36 & 211 \\
NGC 4944 & 7111 & 0.34 & 33.49 & 22.36 & 19.92 & 0.01 & 0.01 & 0.11 & 20.95 & 196 \\
NGC 4952 & 5919 & 0.62 & 12.78 & 10.31 & 18.81 & 0.01 & 0.02 & 0.09 & 19.20 & 232 \\
\cline{1-11} \\
Maffei 1\tablenotemark{*} & \Vhel & 0.73 & 203.60 & 150.17 & 21.25 & \opd & \AI & 0.00 & \SBI & \Vdap \\
\phn & (\Vhelerr) & (0.01) & (10.00) & (7.38) & (0.04) & (\dopd) & (\AIerr) & \nodata & (\SBIerr) & (\Vdaperr) \\

\tablecomments{
(1) Name of galaxy.
(2) Heliocentric radial velocity from SGHb.
(3) Ratio of the semi-minor to the 
semi-major axis, computed from $\log R_{25}$ from RC3.
(4) Effective radius in $I$, corrected 
for seeing effects, from the $r^{1/4}$ 
fit to the surface brightness profile.
(5) Effective aperture radius, computed from $r_e$ 
following \citet{ov81}.
(6) Measured mean effective $I$-band surface 
brightness, defined by SGHa as $<\mu>_e = \mu(r_e)-1.392$.
(7) Optical depth at $1 \, \rm \mu m$ due to dust, 
derived from reddenings of \citet{sfd98} following \citet{mlm02}.
(8) $I$-band Galactic extinction, computed from 
$\tau_1$ following \citet{mlm02}. 
(9) $I$-band correction for K-effect according to 
\citet{mlm02} plus $\rm 2.5 \log (1+z)^4$ correction for 
surface brightness dimming.
(10) Mean effective face-on $I$-band surface 
brightness corrected for seeing, Galactic extinction and 
cosmological effects
(i.e, 
$<\mu>_e^0 = <\mu>_e - A_I^{see} - A_I^{gal} - A_I^c - 2.5 \log b/a$,
where $A_I^{see} = 0.016 \, \rm mag \, arcsec^{-2}$ following SGHa).
(11) Velocity dispersion corrected for aperture 
effect using equation~{(2)} of J$\o$rgensen et al. (1995) 
with $\rm \theta_{obs}=2\arcsec$ (SGHb).
}

\tablenotetext{*}
{
For Maffei~1, the values in columns (3)-(6) 
are taken from the photometry of \citet{bm99}.
Columns (2), (7) and (11) contain the measurements determined in 
this study, with the velocity dispersion corrected 
for aperture effect using equation~{(\ref{eqapcor})}.
Note that column (10) for Maffei~1 does not include
a correction for seeing.
Uncertainties are given in parentheses.
}

\enddata
\end{deluxetable}

\clearpage
\begin{deluxetable}{ll}
\tablecaption{Properties of Maffei~1\label{tabmaf}}
\tabletypesize{\scriptsize}
\tablewidth{0pt}
\tablehead{}
\startdata
Apparent total magnitudes \citep{bm99} & $B_T=13.47 \pm 0.09 \, \rm mag$ \\
\phn & $V_T=11.14 \pm 0.06 \, \rm mag$ \\
\phn & $I_T=8.06 \pm 0.04 \, \rm mag$ \\
\phn & \phn \\
Apparent total colours \citep{bm99} & $(B-V)_T = 2.33 \pm 0.12 \, \rm mag$ \\
\phn & $(V-I)_T=3.08 \pm 0.07 \, \rm mag$ \\
\phn & \phn \\
Heliocentric radial velocity & $v_\odot = \Vhel \pm \Vhelerr \, \kms$ \\
\phn & \phn \\
Velocity dispersion & $\rm \sigma = \Vd \pm \Vderr \, \kms$ \\
\phn & \phn \\
\mg index within $\theta_e/8$ & $\rm Mg_2=$ \mgadopt mag \\
\phn & \phn \\
Intrinsic effective $V-I$ colour & $(V-I)_e^0=$ \colVIe mag \\
\phn & \phn \\
$V-I$ colour excess & $E(V-I)=$ \EVI mag \\
\phn & \phn \\
Optical depth at 1 $\micron$ due to dust & $\tau_1 = \opd \pm \dopd$ \\
\phn & \phn \\
Extinction normalized to $\tau_1$ & $R_B^1=\RB$ \\
\phn & $R_V^1=\RV$ \\
\phn & $R_I^1=\RI$ \\
\phn & $R_{K^\prime}^1=\RKp$ \\
\phn & \phn \\
Galactic extinction & $A_B=$ \AB mag\\
\phn & $A_V=$ \AV mag\\
\phn & $A_I=\AI \pm \AIerr$ mag\\
\phn & $A_{K\prime}=$ \AKp mag\\
\phn & \phn \\
Apparent total magnitudes corrected for Galactic extinction\tablenotemark{a} & $B_T^0 =$ \mBot mag \\
\phn & $V_T^0=$ \mVot mag \\
\phn & $I_T^0=$ \mIot mag \\
\phn & \phn \\
Apparent total colours corrected for Galactic extinction\tablenotemark{a} & $(B-V)_T^0 = $ \colBVt mag \\
\phn & $(V-I)_T^0 = $ \colVIt mag \\
\phn & \phn \\
Semi-major axis of the elliptical isophote at & $D_{25}^0=$ \Do \\
$\mu_B^0=25 \, \rm mag \, arcsec^{-1}$, corrected for Galactic extinction & \phn \\
\phn & \phn \\
Distance\tablenotemark{b} & $D=$ \aveD Mpc \\
\phn & \phn \\
Distance modulus & $\mu=$ \dmod \\
\phn & \phn \\
Absolute magnitudes & $M_B=$ \MB mag \\
\phn & $M_V=$ \MV mag \\
\phn & $M_I=$ \MI mag \\

\tablenotetext{a}
{
Cosmological corrections at this redshift are negligible.
}
\tablenotetext{b}
{
Based on a distance to Coma of 78.8 Mpc,
which is the average value measured from the Fundamental Plane
in Gunn $r$ and from the Tully-Fisher relation in $I$
\citep[see][]{fmg01}, but anchored to the maser distance
of NGC~4258 (7.2 Mpc: see \citealt{hmg99}) rather than
to the Cepheid distance of the LMC (0.05 Mpc: see \citealt{fmg01}).
}
\enddata
\end{deluxetable}
\end{document}